\documentclass{article}
\pdfoutput=1
\usepackage{listings}
\usepackage{PRIMEarxiv}
\usepackage{graphicx}
\graphicspath{{media/}}     % organize your images and other figures under media/ folder

\title{Exploring Security Practices of Smart Contract Developers}
\author{
{Tanusree Sharma, Zhixuan Zhou, Andrew Miller, Yang Wang}\\
University of Illinois at Urbana-Champaign\\
(tsharma6, zz78, soc1024, yvw) @illinois.edu}

\begin{document}
\maketitle

\begin{abstract}
Smart contracts are self-executing programs that run on blockchains (e.g., Ethereum). %Over 100 billions of dollars are controlled by smart contracts, which have become lucrative targets for attacks. 
680 million US dollars worth of digital assets controlled by smart contracts have been hacked or stolen due to various security vulnerabilities in 2021. Although security is a fundamental concern for smart contracts, it is unclear how smart contract developers approach security. To help fill this research gap, we conducted an exploratory qualitative study consisting of a semi-structured interview and a code review task with 29 smart contract developers with diverse backgrounds, including 10 early stage (less than one year of experience) and 19 experienced (2-5 years of experience) smart contract developers.  

Our findings show a wide range of smart contract security perceptions and practices including various tools and resources they used. %Our experienced smart contract developer (2-5 years of experience) participants tended to have more awareness of smart contract security (e.g., vulnerabilities, best practices, resources, tools) and perform much better in the code review task than their early stage counterparts (less than one year of experience).  
Our early stage developer participants had a much lower success rate (15\%) of identifying security vulnerabilities in the code review task than their experienced counterparts (55\%).  
% \yang{add major findings: early vs experienced developers}
%Many participants mainly relied on external audits to ``take care of security.'' They also lacked awareness of relevant information resources and tools. Furthermore, many participants failed to identify common vulnerabilities (e.g., reentrancy, overflow) in code review tasks. In comparison, other participants used a combination of tools, call graphs to ensure security. 
Our hierarchical task analysis of their code reviews implies that just by accessing standard documentation, reference implementations and security tools is not sufficient. Many developers checked those materials or used a security tool but still failed to identify the security issues. 
In addition, several participants pointed out shortcomings of current smart contract security tooling such as its usability. We discuss how future education and tools could better support developers in ensuring smart contract security.

\end{abstract}

% keywords can be removed
\keywords{Smart contract, security, blockchain, developer}

\section{Introduction}
\label{sec1}

%\yang{let's target Oakland 12/2. Submitted papers may include up to 13 pages of text and up to 5 pages for references and appendices, totaling no more than 18 pages.}

%\yang{ccs 1/20. no more than 12 pages long, excluding the bibliography, well-marked appendices, and supplementary material. }

%\yang{abstract + intro 1.5 pages, related work 1.5 pages, method 2.5-3 pages, results 5.5-6 pages, discussion 1.5 pages}

% \yang{frame it more as a smart contract paper rather than DeFi} \tanusree{updated}
Blockchains are cryptographic platforms that can securely host applications and enable the transfer of digital assets in a decentralized manner. Smart contract blockchains like Ethereum are increasingly capable of supporting sophisticated computations (a.k.a., decentralized apps or dApps)~\cite{wood2014ethereum}. Smart contracts are program scripts that define customized functions and rules during transactions, and can run autonomously once deployed on a blockchain~\cite{egbertsen2016replacing}. To support this unique form of computation, domain specific programming languages, such as Solidity and Vyper have been created to allow developers to write smart contracts. 

A wide variety of industry applications in finance, healthcare, and energy have been rapidly exploring the use of blockchain technology and smart contracts to enable system transparency and traceability. Since deployed smart contracts can perform critical functions of holding a considerable amount of digital assets, tokens or currencies in circulation, they become a hotbed for attacks.
% \yang{add some numbers about DeFi, like total value locked (TVL) from defipulse or something like that} \tanusree{updated}. 
%For instance, Decentralized Finance (DeFi) is a booming industry that relies on smart contracts. 
According to DeFi Pulse \footnote{https://defipulse.com/}, there is about \$109B USD worth of total value locked (TVL) as of November 2021 controlled by deployed smart contracts in Decentralized Finance (DeFi) applications. While DeFi is a promising domain and has the potential to disrupt traditional financial systems, there has been several security incidents, in which digital tokens worth of millions of dollars have been stolen. In June 2016, vulnerabilities in the Maker DAO (decentralized autonomous organization) smart contract code were exploited to empty out more than two million Ether, which were worth 40 million USD \cite{sirer2016thoughts}. This attack exploited the reentrancy vulnerability in the `splitDAO' function of the code. Since the code was not designed carefully, a call to the function that behaved as a regular call was modified into a recursive call and used to make multiple withdrawals, effectively depleting the account. Section~\ref{background} provides a technical overview of blockchains and smart contracts.

More recently, since DeFi started skyrocketing in 2020, there has been a new wave of smart contract attacks that led to the loss of hundreds of millions of dollars in value (e.g., \cite{redman_flash_2021}). Clearly, security is critical for smart contracts, and various smart contract security resources and tools have been created. %Prior literature has suggested a few reasons why smart contract security might be difficult, e.g., the need for understanding how blockchain and its various mechanisms (e.g., mining, transactions) work~\cite{coblenz2019smarter}, the open nature of contract code and transactions~\cite{huang2019smart}, and the end-user challenge of understanding key concepts (e.g. private keys) and protecting their crypto assets~\cite{anderson2019money}. 
% {\bf Smart contract developer study}. 
However, it is not clear how people who write smart contracts (we denote as smart contract developers) think about and approach security in smart contract projects. To help bridge this gap, we conducted an exploratory qualitative study consisting of a semi-structured interview and a smart contract code review task with 29 smart contract developers with diverse backgrounds, including 10 early stage (less than one year of experience) smart contract developers and 19 experienced developers (2-5 years of experience). %Since smart contract projects are usually open sourced, code re-use is a common practice and developers often review their own code as well as others' code that they re-use.  

{\noindent \bf Research questions}. Specifically, our study aims to answer the following research questions:
\begin{itemize}
%    \item \textbf{RQ1:} How do smart contract developers think about security of smart contracts?
    %\item \textbf{RQ2}: How do smart contract developers conduct smart contract development?
    \item \textbf{RQ1:} How do smart contract developers approach or ensure security for smart contracts? Are there any differences between early stage and experienced developers?
    \item \textbf{RQ2:} How do smart contract developers conduct code reviews and whether they are able to identify common smart contract security vulnerabilities in the code?
\end{itemize}

    %tackle security challenges in the development process? What are the security practices available (standards, tools, policy) for smart contract development?
    
%\textbf{RQ2}: How smart contract developers tackle security challenges in the development process? 
% (from their development experiences)

%\textbf{RQ3:} What is the level of expertise of developers to identify and solve security problems in smart contract code?
%These vulnerabilities persist despite the rise of security assessment tools. 
{\noindent \bf Summary of findings}. Our study findings show that our participants have a wide variety of smart contract security perceptions and practices, %\yang{make the findings more nuanced by separating early-stage vs experience developers} 
%For instance, none of the early-stage developers (those with less than one year of experience in smart contracts) consider security as an important factor during the system design and development stages, and mainly relied on someone else on the team or external audits to ``take care of security.'' They also lacked awareness of relevant information resources and tools for smart contract security. About half of all participants indicated code re-use as their common practice, especially early stage developers emphasized more on code reuse from open source projects. %Six out of 10 early stage developers indicated code reuse as convenient way to develop smart contract. 
%Furthermore, 13 out of 29 participants (~45\%) failed to identify common smart contract security vulnerabilities such as reentrancy and integer overflow in the code review task. %Eight of them were early stage developers. 
including various tools and resources they used. %Our experienced smart contract developer (2-5 years of experience) participants tended to have more awareness of smart contract security (e.g., vulnerabilities, best practices, resources, tools) and perform much better in the code review task than their early stage counterparts (less than one year of experience).  
In general, the experienced developers tended to have more awareness of smart contract security (e.g., vulnerabilities, best practices, resources, tools) and perform better in the code review task than their early stage counterparts. 
Early stage developers tended to think security is not important for their projects because they do not think their projects have much impact in part because they are often not deployed in the mainnet and thus are not an interesting target for attackers. % whereas experienced developers considered this . %They might be the conception of the importance of security from risk perspective which was negligible for certain project. Besides, 
%This kind of perceptions might result from a lack of security education and awareness of relevant information sources for beginners. 
% 
As a result, they often lack motivations and awareness of smart contract security issues as well as resources and tools about smart contract security. 
In the code review task, our early stage developer participants had a low success rate (15\%) of identifying security vulnerabilities in the reviewed code. %  than their experienced counterparts (55\%).  
% \yang{add major findings: early vs experienced developers}
%Many participants mainly relied on external audits to ``take care of security.'' They also lacked awareness of relevant information resources and tools. Furthermore, many participants failed to identify common vulnerabilities (e.g., reentrancy, overflow) in code review tasks. In comparison, other participants used a combination of tools, call graphs to ensure security. 
%In addition, several participants pointed out many shortcomings of current smart contract security tooling such as its usability. 
% 
In comparison, 55\% of the experienced developers were able to identify these common vulnerabilities and many used a combination of security tools such as static analysis, and interactive call graphs. 
Our hierarchical task analysis~\cite{stanton2006hierarchical} of their code reviews implies that just by accessing standard documentation, reference implementations and security tools is not sufficient. Many developers especially those early staged did check those materials or use a security tool but still failed to identify the security issues.  
In addition, both experienced and early stage developers expressed their challenges with existing smart contract security tools. They pointed out lack of comprehensive tools to identify existing smart contract security vulnerabilities. They also found those tools hard to use (e.g., their user interfaces and the fact that these  are separate tools that they have to use outside of the compiler).

%In this paper, we conduct a qualitative study to understand what types of security practices have been  by smart contract developers, how security fits in the development workflow, etc. We also aim to understand how smart contract developers tackle security risks in the development process. Hence, we conducted experiments (code review tasks) with solidity smart contract developers to identify common security practices and developers' approaches for security improvement in smart contract code. Overall, 

%\zhixuan {add some general findings before contribution} \tanusree{added in abstract. won't it be repetitive?}

{\noindent \bf Main contributions.} Our work makes the following contributions: 
% \yang{major contributions} \tanusree{added}
%\begin{itemize}
%\item 
(1) our rich interview data offer novel results on the various security perceptions and practices of smart contract developers with diverse backgrounds;  %  to investigate security practices in smart contract development across different developer samples: industry developers, researchers, students, freelancers, and DeFi company leads. 
%\item 
(2) results from our smart contract code review task sheds light on how smart contract developers actually examine smart contract code for security vulnerabilities; 
and (3) we discuss implications for security education and tools that could better support developers in ensuring smart contract security. 

% \yang{we might need to move the background section on blockchain and smart contracts here. let's decide after hearing from USENIX}
%We asked developers to perform code review task, showcasing different approaches of solidity developers to identify and solve security vulnerabilities in smart contract code, and identifying challenges that need to be addressed toward a more secure smart contract ecosystem. 
%\item 
%(3) Participants' diverse perceptions of security present the limitations and challenges they face during development process of smart contract. Optimism bias about tackling security within a team and blurred understanding of security appear to be common responses from developers. Part of these prospects are associated with systemic barriers and company culture that needs to be investigated more. 
% however, not as one of some main factors (Gas optimization, functional correctness) they consider. 

% \end{itemize}

%The remainder of this paper is organized as follows. Section 2 gives an overview of the smart contract programming languages and Solidity. Sect. 3 presents and describes Related work in current literature on smart contract and software security, Sect. 4 provides the details of the experiment, Sect. 5 includes  evaluation results; Sect. 6 presents Discussion and  lastly, Sect. 7 reports the conclusion and future work.

\section{Blockchain and Smart Contract}
\label{background}
%\yang{Zhixuan will work on this} \tanusree{updated}

% \review{blockchain as a general concept be explained more briefly with citations to full descriptions for the interested reader, and the background section focuses more on Ethereum and smart contracts.}

The financial industry is seen as a primary user scenario of the blockchain concept. Blockchain is well known not only because of its popular application - cryptocurrency, but also for the characteristics of solving the process inefficiencies in financial services, by tracing ownership over a longer chain of changing buyers overtime.
It is a sequence of blocks, which hold a complete list of tamper-resistant transaction records in a distributed fashion without any central authority \cite{chuen2017handbook, yaga2019blockchain}. Each block contains a timestamp, the hash value of the previous block, and a nonce, which is a random number for verifying the hash using cryptographic means. \cite{chuen2017handbook}. It guarantees reliability and consistency of data by adopting decentralized consensus mechanism \cite{nofer2017blockchain, blakley2021discussion}. Blockchain ecosystem contains multiple components- a core blockchain software that consists of client software such as Go Ethereum \cite{ethereum2017official} in the case of Ethereum. People participating in a blockchain network run this client software, called node, which is often bundled with software/hardware wallets~\cite{amiet2021blockchain}.

%If the majority of nodes in the network agree by a consensus mechanism on the validity of transactions in a block and on the validity of the block itself, the block can be added to the chain \cite{swanson2015consensus}.
%This concept ensures the integrity of the entire blockchain back to the first block (a.k.a. genesis block). Hash values are unique and fraud can be effectively prevented, since changes of a block in the chain would immediately change the respective hash value \cite{nofer2017blockchain}. 

%Being in a decentralized network, it does not need third-party trusted authority. Instead,
%In existing blockchain systems, there are four major consensus mechanisms \cite{zheng2018blockchain}, Proof of Work (PoW), Proof of Stake (PoS), Practical Byzantine Fault Tolerance (PBFT), and Delegated Proof of Stake (DPoS). 

The most popular blockchain systems, Bitcoin and Ethereum, 
%use the PoW mechanism where Ethereum also incorporate PoA (Proof of Authority) \cite{kovan}. We focus on Ethereum in our case for our research on smart contracts. 
Ethereum is a decentralized, open-source blockchain with smart contract functionality. It enables developers to build decentralized applications with built-in economic functions in smart contracts \cite{antonopoulos2018mastering}. While providing high transparency, and neutrality, it also reduces/eliminates censorship, and reduces certain counterparty risks. Ether is the native cryptocurrency of this platform and second-largest cryptocurrency by market capitalization\footnote{https://coinmarketcap.com/}.

Smart contracts are essentially containers of code that encode real world contractual agreements in the digital realm \cite{macrinici2018smart}. It is a protocol that can automatically verify and process the content that represents a binding agreement between two or more parties, where every entity must fulfill their obligations of the contract. Once a smart contract is confirmed by the consensus protocol and submitted to the blockchain, it will be run in terms of the prior negotiation without the interference of any third party. The code execution of smart contract is distributed and verified by the network nodes in a decentralized blockchain network \cite{antonopoulos2018mastering}. It can enable the encoding of complex logic including, payoff schedule, investment assumptions, interest policy, conditional trading directives, pricing dependent on geographic or other environmental input, etc \cite{grech2018madmax}. Gaming, music distribution, remote purchases, and many other business interactions can utilize automatically-enforced smart contracts for secure and verified transactions, without a need for intermediaries or third-party trust \cite{grech2018madmax}.

%Virtually any complex interaction can be captured, with its logic permanently and transparently stored on the blockchain. The possibilities for such logic are endless: it can 
%There are different platforms for smart contracts such as Ethereum, which is the largest and most popular platform for building distributed applications (dApps) \cite{colchester2018blockchain}. 
%Bitcoin which is a enhanced way of capital circulation offering decentralized autonomous market \cite{hsieh2018bitcoin}. 
%Owing to the success in Ethereum \cite{wood2014ethereum}, smart contracts have been widely supported by many blockchains, such as Fabric \cite{androulaki2018hyperledger}, Corda \cite{brown2016corda}, and EOS \cite{xiao2020survey}.

% Smart contract defines a set of rules for governing associated funds, typically written in a Turing-complete programming language \cite{dannen2017bridging} \yang{this is more about smart contracts. try separate the description of smart contracts and solidity}. Contracts can also invoke each other to implement complicated tasks.  For example, on Ethereum platform, it allows its users to run smart contracts on its distributed infrastructure \cite{buterin2014next}. 

%Almost all smart contracts are object-oriented programs. 

There are many smart contract languages, such as Solidity, Serpent, and Vyper. %\andrew{Viper->Vyper}. 
On Ethereum, the most popular smart contract language is Solidity \footnote{https://docs.soliditylang.org/en/v0.8.13/}. Solidity is a JavaScript-like scripting language with static types. 
%It treats contracts as class-like encapsulation constructs, and supports contract inheritance and numerous other features. 
%While Solidity is similar to JavaScript, there are some notable differences. For instance, 
Solidity is strongly-typed and has built-in constructs to interact with the Ethereum platform. Programs written in Solidity are compiled into low-level untyped bytecode to be executed on the Ethereum platform by the Ethereum Virtual Machine (EVM). The EVM is a low-level stack-machine with an instruction set 
%including standard arithmetic instructions, basic cryptography primitives 
for identifying contracts and calling out to different contracts based on signatures, exception-related instructions, and transaction cost computation. Data is stored either on the blockchain, in the form of persistent data structures, or in contract local memory. In Ethereum, solidity smart contracts are generally designed to manipulate and hold funds denominated in Ether.

\section{Related Work}
\label{sec:lit}
% \review{position the work in the literature and justify why this study is needed to complement findings from other developer studies with similar high-level conclusions.}
\subsection{User Studies of Security Practices}
The existing studies on end-users and their attitude towards security and privacy of emerging technologies discussed human factors of software security \cite{green2016developers} and how developers were considered as the ``weakest link."  
%Compared to studies on end-users' security behaviors and perceptions, developers' attitude towards dealing with security has not been investigated proportionately \cite{wurster2008developer}. 
Research highlighted the need for a deeper understanding of the evolving field of software development \cite{pieczul2017developer}. To this end, \cite{acar2016you} outlined an agenda towards understanding developer's attitude, available security development tools, and proposing design suggestions to support developers in building secure applications. There was frequently cited research on developer's lack of security education \cite{oliveira2014s} and security thinking during writing code \cite{cappos2014vulnerabilities}, which was perceived as the reason for different software vulnerabilities. The assumption was that if developers have learned about security, they could better avoid vulnerabilities \cite{wurster2008developer}. To advocate software security education, some presented that there was a lack of security guidelines mandated by the industries \cite{witschey2014technical}, \cite{xie2011programmers} which lead to the lack of ability \cite{oliveira2014s} or proper expertise \cite{baca2009static} to identify vulnerabilities. In this work, we aim to uncover details about security practices in real-world development and deployment in the emerging landscape of smart contracts.

%Researchers explained that security thinking tended to be left out by developers during programming, as vulnerabilities usually existed in corner cases with unusual information flows, and were cognitively demanding to identify \cite{smith2015questions}..

% On that note, however, it is still an open discussion if developers can correlate a particular learned vulnerability or security information with their current working task from their security education.  Oliveira et al. \cite{oliveira2014s} argued that security education are not the root causes of security vulnerabilities. 

\subsection{Smart Contract Vulnerabilities}
%\tanusree{2.2 I was looking to get a better sense of what vulnerabilities were most common and what they were like and how they are handled/fixed/avoided. This might be best done when you introduce the 4 you chose (in 3.3.1). There are hints on those aspects throughout, but a clear recap on how common it seems to be given the literature, how arcane, and how fixed/addressed would be a huge add to the paper.} \update{added literature}

% This section covers the important security attacks and relevant vulnerabilities of smart contract implementation.
% Since smart contracts can hold and manage a large number of virtual currencies, adversaries keep attempting to manipulate the execution of smart contracts to get hold of those. 
Smart contracts are running on distributed and permission-less networks, which leads to vulnerabilities due to the malfunctioning of the smart contract execution. The financial aspect of smart contracts makes them very tempting attack targets, and a successful attack may allow the attacker to directly steal funds from the contracts. Recent literature introduces common vulnerabilities in EVM-based smart contracts, including re-entrancy \cite{perez2021smart, tsankov2018securify, mehar2019understanding}, unhandled exception \cite{brent2018vandal,luu2016making}, integer overflow \cite{lai2020static, perez2021smart}, and unrestricted action or access control \cite{shi2020mechanism, perez2021smart}. For re-entrancy to be exploited, there is a call to an external contract which invokes and re-entrant callback to the contract \cite{mehar2019understanding}. 
%The most famous example of this was the DAO Hack, where \$70million worth of Ether was stolen. The transfer mechanism would transfer ether to external address before updating its internal state and noting that the balance was already transferred \cite{mehar2019understanding}. This gave the attackers a recipe for withdrawing more ether than they were eligible for from the contract via re-entrancy. 
Some low-level operation such as ``send,'' ``transfer,'' and ``call'' are dangerous and can lead to vulnerabilities if they do not handle exceptions \cite{brent2018vandal}. Severe consequences could happen by integer overflow where funds of a contract could become completely frozen \cite{perez2021smart}. Lack of critical authorization check can lead to execution of arbitrary code \cite{shi2020mechanism}.
% Therefore, a vulnerable smart contract is hard to patch and can easily become out of control once deployed. 
% To address this problem, there is a type of contracts that adopt an emergency stop mechanism in some of the functions which is called pausable. It temporarily deactivates those functions and can set the paused state in case of security exploitations. 
%  \yang{there are pausable contracts} \tanusree{mentioned}. 

Blockchain technologies are evolving dramatically which are creating design flaws in smart contract languages. Developers of decentralized apps are often confronted with changing platform features \cite{huang2019smart}. Thus, common software weaknesses such as access control, incorrect calculation, race condition and many other security weaknesses may be amplified on blockchain platforms \footnote{https://swcregistry.io/}. Also, smart contracts are utilized as an anonymous payment method, which is an attractive target for hackers. The infamous disasters involving the DAO \footnote{https://hackingdistributed.com/2017/07/22/deep-dive-parity-bug/} and the Parity Wallet \footnote{https://fortune.com/2016/06/18/blockchain-vc-fund-hacked/4} have highlighted such risks. Attackers exploited programming bugs to steal approximately \$70M USD. Recently, a series of suspicious transactions happened in yCREDIT smart contract where a number of minted tokens were inconsistent \footnote{https://blocksecteam.medium.com/deposit-less-get-more-ycredit-attack-details-f589f71674c3}. xToken suffered attacks with a loss of \$24M USD. One attack was due to a flash loans within the xSNXa contract \footnote{https://blocksecteam.medium.com/deposit-less-get-more-ycredit-attack-details-f589f71674c3}. Recent work shows systemic consensus-layer vulnerabilities due to miner extractable value (MEV) where attackers can front run orders by observing and placing their own orders with higher gas fees~\cite{daian2020flash}.

%\yang{explain front-running as it was brought up in the results. cite the MEV paper.} \tanusree{added}

Recent study identified 16 major vulnerabilities in smart contracts, and limitation of analysis tools to identify the vulnerabilities \cite{praitheeshan2019security}. Most of the known vulnerabilities in smart contracts are related to the fallback function where
% which is an unnamed function triggered when an external caller is sent ETH, the Ethereum token, to the contract, or calls a function that is not defined. When $fallback()$ includes an external function or has potential vulnerabilities, 
attackers could hijack the invoked contract and force it to execute. A study \cite{nikolic2018finding} pointed out that there were 34,200 contracts marked as vulnerable in a million samples due to vulnerabilities related to contract programming languages' design defects. 
% Moreover, since contracts are open source, the attackers can identify the vulnerable spots as well as contract transactions and data may be visible to an adversary. The exposure leads to the fact that smart contract vulnerability is easy to exploit. Although most smart contracts were compiled into bytecodes before deployment, there are various tools that help reverse engineering. Smart contract vulnerabilities may exist in many different layers, including Solidity language, execution logic, and Ethereum Virtual Machine (EVM) design. 
% Nicola Atzei \cite{atzei2017survey} summarized a taxonomy of smart contract vulnerability. It shows that the vulnerable spots can be found through the entire work-flow of smart contract execution. 
% Again in many cases, smart contract developers often do not thoroughly understand the principles of some practices of the blockchain which can lead to security issues in smart contract development \cite{huang2019smart}. \yang{we can cite a few recent attacks to illustrate the problem space} 
Some research focused on demonstrating the difficulties of estimating vulnerabilities in practice \cite{perez2019smart}. 
%suggested that the potential impact of vulnerable code had been greatly exaggerated in
They surveyed six academic projects with 21,270 vulnerable contracts within datalog based queries level exploit discovery in EVM traces to contrast the reported and actual vulnerabilities in those smart contracts \cite{perez2021smart}. 
% The reason for the difference between the exploited and vulnerable contracts is that only less than 10\% of contracts in their study contracts contain Ether, and most of them have a balance lower than 1 ETH. Hence, even if there are vulnerabilities in contracts, those are not exploited since they do not contain ether or economic value. However, if we see development practices in smart contracts, these contracts are often open-sourced \cite{huang2019smart} and can be adopted/borrowed easily for creating a new contract. Therefore, if the new contract has ether in it, then the contract can be exploited due to the borrowed,  vulnerable code.
% \yang{maybe add a recent attack where millions of dollars were stolen.} \tanusree{added}%Therefore, the risk of security issues in smart contracts can be more severe than that in traditional applications.
% We picked some common vulnerabilities and embedded them in sample code during code review tasks, in order to evaluate developers' expertise in identifying security vulnerabilities, either by manual checking, or by using automated security tools.  

\subsection{Smart Contract Security Methods/Tools}
In order to ensure smart contract security, a wide variety of practices have been adopted to operate at different stages of smart contract development life cycle.

\noindent \textbf{Testing tools}. A large number of tools have been developed to analyze either contract source code or its compiled EVM bytecode \cite{perez2019smart}, and look for known security issues, such as re-entrancy, overflow, etc. Some of the well-known tools are Oyente \footnote{https://github.com/enzymefinance/oyente}, ZEUS \cite{kalra2018zeus}, Maian\footnote{https://github.com/ivicanikolicsg/MAIAN}, SmartCheck \cite{tikhomirov2018smartcheck}, ContractFuzzer \cite{jiang2018contractfuzzer}, Vandal \cite{brent2018vandal}, Ethainter \cite{brent2020ethainter}, Securify \cite{tsankov2018securify}, and MadMax \cite{grech2018madmax}. ConsenSys Diligence - an enterprise for comprehensive code reviews of smart contracts developed Mythril \footnote{https://github.com/ConsenSys/mythril} which supports detecting security vulnerabilities in EVM-compatible blockchain and MythX to cover a wider range of security issues \footnote{https://github.com/ConsenSys/mythril}.

% \yang{static and dynamic analysis tools?} \tanusree{yes}

\noindent \textbf{Development and testing frameworks}. Truffle \footnote{https://github.com/trufflesuite} is the most common development framework for smart contracts. It allows developers to write both unit and integration tests. Another development environment is Hardhat which facilitates running tests, automatically checking code for mistakes, and interacting with a smart contract, runs on a development network. It provides plugins for code coverage, measuring gas usage per unit test, automatically verifying contracts on Etherscan, etc. 
% One difficulty of testing on the Ethereum platform is that the EVM does not have a single main entry point and bytecode is executed when fulfilling a transaction. There are mainly two methods used to work around this. The first is to use a private Ethereum network, or a test-net, where it is easy to control the state. The smart contracts are deployed and executed on the private network in the same way they would be deployed on the main Ethereum network. The other approach is to use a standalone implementation of the EVM. 
As a local blcokchain, Ganache \footnote{https://github.com/trufflesuite} is one of the most popular standalone implementation of the EVM built for development purposes. There is also a go-to suite of tools for developers called Remix \footnote{https://github.com/ethereum/remix}. It is known for its convenient browser IDE which supports development, testing and deploying smart contract.

% \yang{this still reads like background. are there existing studies about smart contract developers?} 
\noindent \textbf{Code auditing}. %\tanusree{2.3 Code auditing - given the important of code auditing, it would be important to know if code audits easily find the vulnerabilities you chose. If someone else runs a code audit and takes care of the issues (for example), there's really no need to burden every contract developer with catching the problems earlier. Of course it's the rare kind of vulnerability that's like that, but with DSLs (and coding guidelines?) it may be possible.} \update{updated} 
Code auditing is an integral part of the defensive programming paradigm, which attempts to reduce errors before the software is released. 
%  Also, as smarts contracts can have a high monetary value, 
%Auditing contracts for vulnerabilities is a common industry practice. 
Audits should preferably be performed while contracts are still in the testing phase. However, given the relatively high cost of auditing (usually around 30,000 to 40,000 USD), some companies choose to perform audits later in their development cycle \footnote{https://solidified.io/}. 
%As system developers and operators are gradually aware of the importance of blockchain security, more and more auditing tools have emerged. 
Some of the tools used frequently in auditing are Surya \footnote{https://github.com/ConsenSys/surya}, Mythril \footnote{https://github.com/ConsenSys/mythril}, and MythX.Third party companies who perform audits include Trail of Bits, OpenZeppelin, ConsenSys Diligence, etc. 
% Different tools may have different advantages including degree of automation, accuracy and efficiency. 
During audit, vulnerabilities are detected in three main ways: (1) extracting features of malicious code, and doing semantic matching on source code; (2) using mathematical approaches to prove a system's completeness, where auditor specifies every possible input and lists every situation that might happen; (3) generating a control flow graph by contract's logic units by which auditor can traverse all code paths 
%to reveal how the variables are passed through the program in order 
to detect logical design flaws. 
%However, Solidity is a high-level programming language with many potentially vulnerable functions. 
%Thus auditing cannot capture and eliminate all security vulnerabilities for different edge cases and unknown vulnerabilities which requires lower-level inspection of execution traces \cite{perez2019smart}. 
Furthermore, code auditing depends on many factors such as time, budget, and resources availability in different organizations. Code auditing is an extra layer of security instead of a primary consideration. Recent research has been driving towards security from the beginning posture rather than depending solely on auditing \cite{puhakainen2010improving,assal2019think}

% \zhixuan{is (2) static analysis and (3) dynamic analysis?} \tanusree{it can be both. also can be manual}

\noindent \textbf{Bug bounty programs}. Another common practice for organizations is to run bounty programs. It remains ongoing throughout a contract's lifetime and allows community members to be rewarded for reporting vulnerabilities. Companies or projects run bounty programs, e,g., the $0x$ project \footnote{https://0x.org/docs/guides}offers bounties as high as \$100K USD for critical vulnerabilities.

% \yang{how much do they offer?} \tanusree{its not specific. just mentioned that "as high as \$100,000 for critical bugs"}, 

%In this paper, we will present how smart contract developers leverage and perceive helpfulness and usability of security tools, as well as report how they follow existing guidelines, best practices, and standards.

\subsection{User Studies of (Secure) Smart Contract Development}
Recent work has focused on security assessment tools for smart contracts \cite{destefanis2018smart} and comparing existing tools for their capabilities to identify vulnerabilities\cite{mense2018security, brent2020ethainter}. The majority of the tools were still in progress, and several shortcomings were identified for improvement \cite{dika2017ethereum}, such as the user interfaces for visualizing results. Also, the root causes for the occurrence of severe smart contract vulnerabilities were discussed \cite{li2018survey}, but notably only technical factors were covered. %; but this is actually insufficient in any environment with human involvement. 
These approaches were helpful, however they did not address developers' actual practices with the underlying programming languages which could facilitate writing buggy code in the first place. A closer look at the incidents revealed that eventually a large number of vulnerabilities in smart contracts occurred because of developers' mistakes \cite{li2018survey}.
% Furthermore, taking into account the immutability of a blockchain,  classical mitigation of vulnerabilities is actually not enough. 
Hence, avoidance of vulnerabilities in smart contracts meant secure development with users in mind during the development life cycle \cite{destefanis2018smart}. This included a well thought-out design as well as adhering to best-practice patterns for ensuring security \cite{destefanis2018smart}, and involving users/developers during the design process. Research proposed that developers' needs should be considered as a primary requirement in new tool design for smart contracts \cite{coblenz2019smarter}. 
% Facilitating safe development by detecting relevant classes of serious bugs at compile time \cite{praitheeshan2019security} and
%Research also suggested leveraging the properties in common blockchain environments to improve safety and developers' effectiveness \cite{praitheeshan2019security}.

There is a lack of empirical user studies that examine smart contract development, particularly security practices. One exception is a study by Parizi et. al~\cite{parizi2018smart} where they compared the uses of smart contract languages: Solidity \footnote{https://solidity.readthedocs.io/en/develop/}, Pact \footnote{https://pact.kadena.io} and Liquidity \footnote{https://github.com/OCamlPro/liquidity}. This study is focused on the usability aspect of smart contract programming languages by measuring the amount of time for novice participants to complete programming tasks. 
%However, this is only one aspect of usability \cite{parizi2018smart}. 
To the best of our knowledge, our study is one of the first to empirically explore smart contract developers' security perceptions and practices. Despite the wide range of vulnerabilities and exploitation reported in smart contracts \cite{perez2019smart}, and existing research on assessing the effectiveness of security analysis tools \cite{sayeed2020smart}, we still know little about whether, when and how smart contract developers deal with the security aspect of their contract code. We try to bridge these research gaps in this work. 

% Considering the substantial financial value associated with smart contracts, they should be tested and analyzed before deployment. There are several challenges in smart contract development using Solidity language including, lack of knowledge among developers about the usage and implementation of smart contracts since the technology is still in an early stage \cite{wohrer2018smart}; limitation of defined best practices for the programming and testing methods \cite{parizi2018smart}; lack of usable tools for errors detection \cite{staples2017risks}, \cite{destefanis2018smart}.

% \yang{articulate a couple of main research questions.} \tanusree{added}

% To address the problems in existing smart contract research space, we seek to explore current practices in smart contract development and developer's understanding and awareness of security in smart contract. Specifically, we propose the following research questions:

% \textbf{RQ1:} What are the security practices available (standards, tools, policy) for smart contract development?

% \textbf{RQ2}: How smart contract developers tackle security risks in the development process? (From their development experiences)

% \textbf{RQ3:} What is the level of expertise of developers to identify and solve security problems in smart contract code?
\section{Method}
\label{sec:method}

%\yang{Tanusree will work on this}

Our study was inspired by user studies of software developers' security practices (e.g., \cite{gorski2018developers,parizi2018smart,assal2018security}). We conducted a user study with smart contract developers. The study included a semi-structured interview to understand participants' smart contract development experiences and practices particularly around the security aspect. The study also includes a smart contract code review task to understand how developers identify vulnerabilities in smart contracts. Since Solidity is the most popular language for smart contracts \cite{dannen2017bridging}, and has a simple syntax~\cite{parizi2018smart}, we decided to focus on Solidity developers in our study. 
%We recruited our participants via various channels such as Twitter, LinkedIn, developer forums, and GitHub (e.g., those worked on Solidity projects before).
%
%\yang{briefly talk about the screening survey} \tanusree{in section 4.3, screening survey is briefly discussed. Should these be added here too?}
%In the initial interview component, participants were asked questions about their current practices, experiences and challenges in developing solidity smart contracts, particularly their security. In the subsequent study component, participants were also asked to do a code review of one of two smart contracts that we prepared (more details below). Lastly, participants completed an exit interview by providing suggestions for expected features of smart contract security tools as well as feedback on our study. %
%Two researchers coded participants' interview responses. They also evaluate the code review data carefully to arrive at conclusions. 
%
The study was approved by the IRB %the Ethics Review Board of University of Illinois at Urbana Champaign. \yang{it's blind review so don't reveal our identities and affiliations}
%Before user experiment with each participant, we explicitly stated our identity and research intentions. 
and conducted online during the summer of 2021. Each study session took about one hour and each participant was compensated a gift card worth of \$30 US dollars. The study data were collected upon participants' permission. All quotes included in this paper have been anonymized to protect privacy of the participants. 
%We conducted a pilot study with four smart contract developers to test our study design including interview questions and code review tasks. We modified our code review contracts accordingly based on the suggestions from pilot study. 
Below we describe the protocol of our study.

\subsection{Participant Recruitment}

% \review{I think you explain clearly how you did participant recruitment. I think however, that you should clearly explain that that is part of your larger study. I recommend having a section that states that it is part of your formal study and within that have a subsection on recruitment process.}

% \update{moved this section here based on IEEE paper structure}

%Recruiting smart contract developers was a challenge for our research team. Usually these developers are very busy with DeFi projects, which are often lucrative. For instance, smart contract developers were offered \$100 to answer a short survey about a commercial smart contract tool in a recent DeFi hackathon. We as academic researchers will not be able to offer that level of monetary compensation. 
To help reach a wide range of developers, we recruited through different methods: (1) snowball sampling from our contacts in the broader Ethereum community, (2) posting on our Twitter and Facebook as well as $ethresear.ch$ and Discord channels, and (3) contacting Solidity developers of public smart contract projects on GitHub. 

%For the first method, participants were invited to participate in our research based on their availability and consent. We carefully read through the responses of the screening survey to measure the eligibility for participant inclusion. 
We selected participants based on the responses of our screening survey. Respondents were invited to our study if they met our selection criteria: a) is a solidity-based smart contract developer; b) has some smart contract development experience; %(< 1 year - 5+ years) \yang{what's minimum requirement?} \tanusree{familiarity with solidity smart contract development}
 c) should provide proper explanation of their developed smart contract(s). 
We did two waves of recruitment. In total, we received 67 responses from our screening survey. We reached out to 38 of them via email based on our selection criteria. In total, 29 people agreed to participate in our study. %\tanusree{What was the distribution of participants across each recruitment method? Do you have any indications of how "spread out" participants were?}
Eight participants were from word-of-mouth by our contacts in the Ethereum community. Two were from GitHub public projects. 19 were from our postings on Twitter, Facebook, Discord, and the ethresear.ch online forum. %\yang{provide descriptive stats on age and gender, like mean and standard deviation} %We were able to recruit 2 more participants from Github smart contract committers. 
We decided not to recruit further because no new insights emerged from our four latest study sessions, indicating a sign of theoretical saturation~\cite{merriam2015qualitative}.

The participation in our study was completely voluntary, and participants were allowed to quit at any time. Each  participant received a \$30 Amazon e-gift card upon completion of the study. %Compensation was be prorated in a way that recognizes time and effort put in. 
The whole study lasted about one hour, where the initial interview took about 25 minutes, the code review task 25 minutes, and the exit interview about 10 minutes.

\subsection{Initial Interviews}
Since we did not know much about the phenomenon under investigation (security practices of smart contract developers), we felt that an exploratory qualitative study would be appropriate since it will provide rich data about the phenomenon. Therefore, we started with an semi-structured interview. In the study, we first presented the study consent to our participants. Once they have agreed, we proceeded with a semi-structured interview. We designed the interview scripts based on our research questions outlined in the introduction section. The entire script can be found in our study GitHub\footnote{Our study repo includes study scripts, smart contracts for the code review task, analysis code book:  https://github.com/AccountProject/SmartContract}.
% \yang{can we put the interview script and the code review task description in the Appendix. if we don't have space in appendix then upload it to an anonymous account and put the link here.} \tanusree{I will put all script, task description, codes in githublink}
% 
We started by asking about their general background including their usage of programming languages, their role and experiences in developing software, as well as their experience and motivation on how they started smart contract development. Sample questions are like ``\textit{What role do you typically play in cryptocurrency projects?'', ``\textit{What programming languages do you use regularly?}'', ``\textit{ How did you learn smart contract development?.}''} Then we asked a series of questions to understand their knowledge and experiences of developing smart contracts and handling security issues thereof. Their experiences include tools for writing contracts, guidelines for smart contract development, factors considered and challenges encountered during development.

To understand how smart contract developers tackle potential security risks in the development process, we also asked about their current practices (e.g., use of coding standards, policies, and security analysis tools as well as any educational resources) related to smart contract security. %This provided us the knowledge of their current method and tools of smart contract security assessment, code review process. For deeper understanding of their everyday work on smart contract development and security handling process, 
To get rich anecdotal data, we also asked about their personal experiences and stories of how they handled specific smart contract security related issues in the past. %Lastly, we asked them their sources of learning on smart contract security and development.

Thus, interview questions get at not only general practices but also specific cases (e.g., \textit{``What are the most frequent security issues you have encountered? Please tell us about any incident that you can remember when your developed smart contract was exposed to certain security vulnerabilities. What types of methods/tools are you/your organization currently using to ensure the security of smart contracts? Have you used any automated security analysis tools? Can you explain more about the tools? Is it easy or hard to use?''}). Through qualitative analysis of interview data, we identified coding themes emerged from the data on development practices of smart contract developers. Example themes include \textit{``functionality correctness is a priority,''} \textit{``security is not  important when projects are deployed on testnets,''} \textit{``Solidity language limitations make smart contract security hard.''} The code review task was designed to show how ``concretely'' they go about reviewing the code and identifying potential vulnerabilities.

\subsection{Smart Contract Code Review Task}
%\tanusree{Why were participants limited to 25 minutes for the code review? Was this based on common practices in smart contract development?} \update{done}

%\tanusree{Why was a well-known piece of code (i.e., the "hello world" smart contract equivalent) reasonably complicated to avoid participants simply identifying any added code as the vulnerability.} \update{done}

The next study component was a smart contract code review task. Since code review is a common task in smart contract development, this task was designed to understand how smart contract developers conduct code review particularly for identifying security vulnerabilities. Each participant was asked to review one smart contract that we created. %Through this task, we observe different approach for developers to solve the problems as well as the level of expertise in addressing security vulnerabilities in smart contract development. 
Specifically, they were instructed to do the following: 
%\begin{itemize}
(1) share their computer screen and allow us to record the screen with their permission to understand how they conduct the code review; 
(2) have at most 25 minutes for the code review; 
(3) search/use any resources/tools they need; 
(4) i. review the code; ii. identify security vulnerabilities and/or areas for improvement; iii. modify the code accordingly; %\tanusree{3.3 "modify the code accordingly" - that's not really a code review task then, in the classic sense. I wonder how likely participants were to not find anything they couldn't fix. In other words, I wonder if just a code review would have had slightly different outcomes.}
(5) explain the modifications and rationale behind them.
%\end{itemize}

%After finishing the tasks, we conduct exit interview with the participants to ask general questions about awareness and understanding of security practices in smart contract development as well as their feedback on our study.

\subsubsection{Code Review Task Design}
\label{sec:task-design}
%We designed tasks that were short enough so that the uncompensated participants would be likely to complete them before losing interest, but still complex enough to be interesting and allow for some mistakes. 
We designed this study component to model the real-world code review task that developers would be reasonably expected to encounter in their smart contract development. We chose to include two vulnerabilities in each smart contract where one vulnerability is more well-known and should be easier to identify than the other. We measured the difficulty level of these vulnerabilities based on our pilot participants' feedback. We also included some minor code writing standard issues, such as indentation, space/tabs, and blank lines which should be easily detectable. 

We chose four smart contract vulnerabilities based on the recent literature on smart contract exploitation (\cite{perez2021smart,sayeed2020smart}) and reports \footnote{https://swcregistry.io/} on most frequent exploitation. The four chosen vulnerabilities are: 1) re-entrancy, 2) unchecked low level calls, 3) integer overflow, and 4) improper access control. The first two vulnerabilities were included in one contract and the last two vulnerabilities in another contract. The contract code can be found in our study GitHub %\href{https://github.com/AccountProject/SmartContract}{GitHub} 
repo. %The four vulnerabilities and their effective mitigation strategies are summarized in Table~\ref{Tab:tab-vul} (Appendix).
Table~\ref{Tab:tab-vul} summarizes the four security vulnerabilities we included in the code review task  and ways to avoid/address these vulnerabilities. 
\label{tables}

\begin{table*}[!t]
    \centering
    \caption{Vulnerabilities in code review task as well as effective strategies to address these vulnerabilities.}

    \small
    % \begin{{tabular}{p{0.11\linewidth} | p{0.335\linewidth}| p{0.50\linewidth}}
    %\begin{tabular}{*{3}{p{.39\linewidth}}}
    \begin{tabular}{p{.2\linewidth}|p{.4\linewidth}|p{.3\linewidth}}
     \hline
   Vulnerability &  Description & Possible Prevention Technique \\
   \hline
      Reentrancy & Calling external contracts is that they can take over the control flow, and make changes to the data that the calling function is not expecting. & Make sure it does not call an external function or use the Checks-Effects-Interactions pattern. \\
     \hline
      Unchecked Low-Level Calls & This can lead to unexpected behaviour and break the program logic. A failed call can even be caused by an attacker, who may be able to further exploit the application & Ensure to handle the possibility that the call will fail by checking the return value.\\
     \hline
      Integer Overflow & An integer overflow occurs when an arithmetic operation attempts to create a numeric value that is outside of the range & Use of vetted safe math libraries for arithmetic operations consistently throughout the smart contract system.\\
      \hline
      Improper Access Control & Exposing initialization functions by wrongly naming a function intended to be a constructor, the constructor code ends up in the runtime byte code and can be called by anyone to re-initialize the contract. &Implement controls so withdrawals can only be triggered by authorized parties or according to the specs of the smart contract system.\\
      
      %\midrule
%       \textbf{Functional Issues} &  & \\
% Safe Math & Arithmetic operations in Solidity wrap\\
      \hline
    \end{tabular}
    \label{Tab:tab-vul}
  \end{table*}

We created our two smart contracts using a boilerplate contract of ERC-20 token which provides basic functionalities to transfer tokens and allow tokens to be approved so they can be spent by another account. The ERC20 token contract is commonly used as a basis for smart contract (DeFi) projects. Numerous DeFi projects and open source smart contracts adopt the ERC20 standard interface. The ERC20 standard interface can create tokens on Ethereum and can be re-used by applications such as wallets and decentralized exchanges. We believed that most smart contract developers would have some familiarity with this contract and our pilot study confirmed our assumption. Our contracts for code review have the same basic ERC-20 functionalities.

To select smart contracts vulnerabilities, we conducted an extensive review of exploited smart contracts from different security vulnerability reports, such as Smart Contract Weakness Classification and Test Cases (SWC) \footnote{https://swcregistry.io/}, Consensys Known Attacks \footnote{$https://consensys.github.io/smart-contract-best-practices/known_attacks/$}, different GitHub repositories and $etherscan$ source code of abandoned or previously exploited smart contracts \footnote{$https://etherscan.io/directory/Smart_Contracts$}. To make our code review task realistic for participants to complete in $25$ minutes, we trimmed down some auxiliary functions from the selected contract sample code. Both of the contracts for code review are ERC-20 based and there are no major differences in size. We then modified the contract by adding the vulnerabilities.  %(Figure~\ref{fig:code-snippet-new}). 
We got feedback from solidity programmers and tested the contracts in our pilot. We included comments in the contract code for participants to understand the overall context of the contract functions. We also added explicit comments to those function that are not part of the ERC20 token standard (interface). %For instances, in one contract, we added $deposit()$ and $withdraw()$ functions. Therefore, we added an explicit comment saying that "/* Functions below are specific to this sample token and * not part of the ERC-20 standard */".  Our created smart contracts were in Solidity version 0.7.0 or higher. %This is because the default Solidity version in Remix, a popular smart contract development IDE, at the time of our study (March-May 2021) was 0.7.0. Each contract has around 80-100 lines of code excluding comments.  

During the task, we provided the instructions verbally. After giving the instructions, we provided the GitHub link which contained the smart contract for review to the participants through Zoom chat. We provided some extra time for participants to set up their development environment before we started counting the 25-min task time. This set-up time which was not counted in the 25 minutes. Since the target participants are very busy, we limited our study to about one hour. Code review time (25 minutes) is based on the code length, which varies significantly in practice (10-1000 lines of code). It is a limitation but for practical reasons it was a trade-off between study time and participant's availability.   %During this task, participants shared their screen and allowed us to recording their code review activity and the resources they searched during the task. 

%After completion time, participants were asked to explain their modification/comments on the area of improvements they made during the task and the rationale behind those suggested improvement. 

\subsection{Exit Interview}
Once the task was completed or the time ran out, the participants were asked a few questions in the exit interview. We asked their opinions about the task they worked on and how they perceived the difficulty of the task. We also asked their overall experience with the code review task and if they would like to share any other experiences of security practices in smart contract development. We also asked if they had any feedback on this study and whether this study has changed their perspectives and/or attitudes towards smart contract security. Finally, we asked about their desired features of security tools to help them better ensure smart contract security. We received the modified code by each participant via email. This data allowed us to determine whether they correctly identified the vulnerabilities.

\subsection{Pilot study}
%\yang{briefly explain and refer to the section in Appendix}

We conducted two rounds of pilot study with a total of four smart contract developers to test our study design. We revised our interview questions and code review tasks based on their feedback.  
%In the first round, we had participants who were doctoral student researchers in the blockchain and smart contract security area. Participants suggested that contracts in the code review task should be more realistic by having common functionalities (e.g., ERC20). We modified our code review contracts accordingly. To test the updated study materials, we had a second round with another two participants who were blockchain researchers/developers. They made suggestions to add questions in interview to understand what role(s) the participants played in their smart contract projects. We modified our interview questions accordingly. 
Details of the pilot study and the changes we made based on the pilot results are described in the Appendix~\ref{appendix:pilot}.

\subsection{Data Collection and Analysis}
Study data were obtained primarily through Zoom audio/video/screen recording of the interviews and the code review task upon participants' permission. We collected each participant's interview responses, smart contract code from the code review task, think-aloud responses, and exit interview responses during the session. %Data collection also includes recorded video of screen sharing of participants' code review task, information of resources searched in online during the task.
We transcribed the recordings and analyzed the transcripts using thematic analysis~\cite{boyatzis_transforming_1998}. 
%
%Interview results reported in this paper represent participants’ self explained responses on smart contract security practices and experiences. Data analysis for those responses followed an inductive approach. 
% 

%\tanusree{Why was a Cohen's Kappa of 0.68 used? What is the argument for this threshold? - will update with new inter coding score-E}
%\tanusree{inter-coder agreement rate was on the low side. It would be important to have some explanation as to why this happened and if possible go back and re-code the data. Providing an explanation as to why the inter-codder agreement was low and if possible (partially)re-code the data. -C}

%\yang{update with the new inter-coder rate} \update{done}

\subsubsection{Qualitative Data analysis}
Two researchers performed open coding independently on a sample of the data (20\%). Then they met regularly to discuss the coding and reached a consensus on a shared code book before coding the remaining data. We calculated the inter-coder reliability in Cohen's Kappa, which was 0.94 and considered very good~\cite{mchugh2012interrater}.  %$0.81-1.00$ represent almost perfect agreement. 

%In our open coding of the data,  coded interview data into 479 unique excerpts.
Our open coding followed an inductive analysis method to explore practices and behaviours towards smart contract development. Our codebook includes 46 codes. 
%{\color{red} \bf@Tanusree: remove the current code book from the study github repo because it contains raw data. just upload the code book that contains the codes not the raw quotes} \tanusree{deleted. will add the code book}
Then data and concepts that belonged together were grouped into sub-categories. Further abstraction of the data was performed by grouping sub-categories into generic categories, and those into main categories. We  then grouped related codes, organized them in high-level themes, and iterated this process to finally produce 18 themes to interpret the results of smart contract developers' security practices, security concerns and individual experiences and challenges with tools and development methods. Some example themes are: perceived priorities in smart contract development, use of information resources, use of development tools, use of smart contract security tools, and limitations of smart contract security tools. 

%Other combination may be possible, but we found this list ~Table\ref{} helpful to interpret our study results.}

%\tanusree{a table in appendix with emerging theme in 4 main categories and sub categories} \tanusree{Do we need to add all the themes in appendix for thus venue?}

\subsubsection{Assessing The Code Review Task Outcome} For the code review task,
% {\bf Identifying and fixing vulnerabilities.} 
recorded videos of the participants performing the code review task were analyzed to measure the success rate, based on whether a participant correctly identified the security vulnerabilities in the smart contract. 
%\update{To provide task success score, we assigned participants a score of 1 (if participant successfully identify security vulnerability and provided solution/comment for assigned task), or 0 (otherwise). %Table ~\ref{Tab:exp} includes the task score for each participant.  %We also analyzed whether the participant mistakenly identified vulnerabilities that were not present (i.e., false positives). 
Prior to conducting the lab study, we created and verified the correct/secure solutions for each task. The general ideas for these solutions are described in Table~\ref{Tab:tab-vul}. This ensured that we could verify whether participants successfully identify the vulnerabilities and provide a correct/secure answer or fix. We also paid attention to how they conducted the code review, e.g., whether and how they searched and used any resources/tools. Specifically, we conducted a  hierarchical task analysis~\cite{stanton2006hierarchical}, a common Human-Computer Interaction (HCI) technique, %\yang{cite}
to break down the process of how a participant performed the code review into detailed steps and to help identify how the processes of successful and failed code reviews differ. We also checked whether their suggestions or modified code could potentially fix the vulnerabilities.

\section{Findings}
%\tanusree{Reframing the paper so that it clearly portrays the data collected. Provide a deeper analysis that takes advantage of the richness of the underlying data.}
%\tanusree{provided the frequency of some of the themes when reporting them. Please consistently do this for all the qualitative findings. Provide the frequency of the themes consistently throughout the results section.}
In this section, we present our study results
in diverse views and practices of smart contract security. We have found none of the early stage developers consider security as a main factor and lack awareness of relevant relevant information resources and tools for smart contract security. We also discuss our lab study in terms of task success rate and time to identify security vulnerabilities, use of different code review approach. We found that experienced developers (3-5 years of experience) were able to identify these common vulnerabilities and many used a combination of security tools such as static analysis, and interactive call graphs.

%\yang{add high-level summary of results: diverse views and practices on smart contract security} \tanusree{added}

\label{sec:findings}

\begin{table*}[!t]
  \centering
%   \resizebox{\columnwidth}{!}
  \resizebox{1.0\linewidth}{!}
  {\begin{tabular}{l l l l l l l l l }
    % \toprule
    {\small\textit{ID}}
    & {\small \textit{Gender}}
    & {\small \textit{Country}}
    & {\small \textit{Occupation}}
    & {\small \textit{Year(s) Exp.}}
       & {\small \textit{Focus}}
    & {\small \textit{Types of Projects Developed}}
    \\
    \hline
   \textbf{Early-Stage}\\
    
    P1 & Male & USA & Developer, DeFi Company& $<$1 & Backend& Distributed systems, cryptography\\
    P3 & Male & USA & Masters Student, CSE & $<$1 & Backend & Security applications in Software\\
    P4 & Male & Australia & Researcher, DeFi Company & $<$1 & R\&D & Smart Contract \\
    P5 & Male & India & Bachelors student, CSE & $<$1 & Front-end & React apps development\\
    P7 & Female & USA & Masters Student, CSE & $<$1 & Backend & Security product on authorization \\
    P9 & Male & USA & Professor, CSE & $<$1 & Backend & Security auditing of smart contracts \\
    P16 & Male & India &Freelance Smart Contract Dev  & $<$1& Backend&blockchain based project   \\
    P20 & Male & USA &Bachelor Student, CSE & $<$1  & Backend &   Class project on Blockchain\\
    P24 & Female&Canada  &PhD Student, CSE & $<$1 & Full-Stack & Research on blockchain \\
    P28 & Male & USA & Developer, De-Fi Industry & $<$1 &  Backend & Distributed Ledger Technology\\
      \hline
  \textbf{Experienced} \\
    P2 & Female & USA & Developer, DeFi Company  & 2-3 & Front end & Software for Retail\\    
    P6 & Male & USA & PhD Student, CSE & 3-5 & Backend & Formal framework in cryptography\\
    P8 & Male & Ghana & Freelance Smart Contract Dev & 3-5 & Full-stack & Blockchain software from scratch \\
    P10 & Male & USA & Developer, DeFi Company & 3-5 & Full-stack & Software for privacy, blockchain \\
    P11 & Female & China & Dev, DeFi Company & 2-3 & Full-stack & Application for finance, game, NFT \\
    P12 & Female & Egypt & Freelance Smart Contract Dev & 3-5 & Full-Stack & Blockchain based project  \\
    P13 & Male & Iran & Software Developer & 3-5 &  Backend & Application for health sector \\
    P14 & Male & UK & Co-Founder,dev, DeFi Company & 2-3 & Backend & Developed DAO, solidity projects  \\
    P15 & Male & India & Software Developer& 2-3 & Backend & Machine learning and AI project\\
    P17 & Male & India & Software Developer& 2-3 & Backend& Smart contract for donations   \\
    P18 & Male & USA& Dev, Defi Company & 2-3 & Full-stack&Research and architecture  in DeFi \\
    P19 & Male & USA & CTO, DeFi Company  & 3-5 &  Backend &  Snapchat and DeFi token  \\
    P21 & Male & India &Bachelor Student & 2-3 & Backend & Hyperledger and Ethereum related project\\
     P22 &Male & Germany  &PhD Student, CSE & 3-5  & Backend & Blockchain for data storage\\
     P23 & Male& Australia & Developer, DeFi Industry&  2-3 &Backend & Financial smart contracts (Tracer)\\
     P25 & Male & Greece & Developer and Researcher & 2-3 & Backend & Gaming / betting smart contract\\
     P26& Male& Germany & Developer, DeFi Industry & 2-3 & Backend & Auditing Smart Contract\\
     P27 & Male & Canada & Developer, DeFi Industry & 3-5 & Backend& Layer 2 project in smart contract\\
     P29 & Male & New Zealand & Developer, DeFi Industry & 3-5 & Backend& De-Fi Project\\
     \hline
    % \bottomrule
  \end{tabular}}
  \caption{Participant demographics and background. We consider those with less than one year of experience in Solidity as early-stage developers and those with longer experience as experienced developers.}
  \label{Tab:Tab1}
\end{table*}
%\yang{Yang will work on this}
% 

%\subsection{Interview Findings}

\subsection{Participants} 
We had a total of 29 participants. 24 of them are male and the rest are female. 
 %\yang{do we have age?} \tanusree{we don't have age info} 
%\update{Nine had less than one year of experience in Solidity and thus we considered them as early-stage developers. The other 15 participants had longer experience (2-5 years) and we considered them as experienced developers.} 
More than one third of our participants are from the US, and the rest are from many other countries, including India, China, Australia, Ghana, Egypt, Iran, UK, Canada, Germany, New Zealand, Greece. Our participants include fourteen smart contract practitioners who work full-time in the DeFi industry, three freelance smart contract developers, three software developers mainly worked in other domains, eight college/graduate students, and one professor. %Their years of smart contract development experience ranged from less than a year to five years.
Table~\ref{Tab:Tab1} summarizes our participant demographics. 

To explore the relationship between our participants' different levels of Solidity development experiences and their security practices, we categorized our participants into two groups: {\it early-stage} (less than 1 year of experience) and {\it experienced} (2-5 years of experience) in Solidity smart contract development. 
We had 10 early-stage developer participants and 19 experienced developer participants. 

\subsection{Programming and Development Background}
%\textbf{RQ1: What are the security practices available (standards, tools, policy) for smart contract development?}

\noindent \textbf{Programming background}. %Within participants' response during interview, we acquired the details of their programming language use on daily basis. 
% \yang{would be nice to separate early-stage vs experienced} \tanusree{done for now. will update again after interviews}
%We started with learning about our participants' programming background. Many
Our participants reported using multiple programming languages. Besides Solidity, other languages mentioned are Python (55\%), Javascript (39\%), C++ (24\%), Java (13\%), and C (13\%).  %(10\tanusree{will update final number/percentage later}), Javascript (7), C++ (6), Java (3), C (2). %There includes other languages, including, Perl, Golang, C#, Rust, Haskell, Typescript, R, PHP, Matlab, SQL, .Net. % 
%
%Early-stage developers were frequently experienced in Python, Javascript, C++, SQL in addition to Solidity. On the other hand, experienced developers frequently had programming background in Python, Javascript, C++, Typescripts, etc.
%
 %Six out of 16 participants had direct experience with security/ software security. Some of the projects mentioned by participants are: distributed system and cryptography; Security feature development in products; Formal framework in cryptography; authorization security in product; Security auditing in smart contract; privacy for software application and blockchain. Details are presented in ~\ref{Tab:Tab1}
%
Besides smart contract development, they also reported other expertise such as backend development (69\%), front end development (7\%), full-stack (21\%) and research (4\%). 
%Figure~\ref{fig:prog-background} (Appendix~\ref{app:Extra-Figures}) summarizes participants' programming background.

%Seven out of ten early-stage developers' main application development area was Backend, one had expertise in Front-end, one was full stack developer, and one had expertise in smart contract research. In contrast, 13 out of 19 experienced developers' main area of expertise was Backend, five of them were full-stack developers, and one took the main role as a Front-end developer.

\noindent \textbf{Information resources used for smart contract development}. 
Our participants mentioned a number of information resources that they used for smart contract development. Common sources are Solidity documentation (34\%), Google search (31\%), Stack Overflow (17\%), YouTube (14\%), as well as documents published by the Ethereum Foundation (14\%), OpenZeppelin \footnote{https://openzeppelin.com/} (14\%), and ConsenSys \footnote{https://consensys.net/}(14\%). %There are also resources, such as, Solidity Wiki,  GitHub issues, solidity site, books,  Teammate, and Online Courses. 

\noindent \textbf{Tools for smart contract development}.  
Participants mentioned a large set of tools they used to develop, deploy, and test their smart contracts. Most frequently used tools included Truffle (76\%), Remix (55\%), Ganache (38\%), HardHat(34\%), and Waffle (21\%). %Besides these most common tools, they also mentioned a number of plugins, editors, packages, libraries and APIs, including VS code, chaiMocha packages, MythX, React, IPFS, infura, tenderly, etc. 
\noindent \textbf{Practices of code re-use}.
%\tanusree{Code reuse (many participants reported doing that, and we will add a new subsection in results)} \update{done}
%Since smart contract projects are usually open sourced, code re-use is a common practice. % and developers often review their own code as well as others' code that they re-use.  
%Code reuse tends to be at the center of maturity for several programming languages as the community builds increasingly powerful code libraries and frameworks on top of previous work. For instance, Perl has CPAN, Ruby has RubyGems, and Javascript has npm. It’s clear that code reuse in the Solidity ecosystem is still in its infancy. 
%In Solidity, there are different ways that code can be reused. First, Inheritance happens when a contract's source code is copied into the code of a new contract \cite{chen2021understanding}. Inheritance typically happens from abstract contracts. Another type of reuse is represented by library function calls. Libraries are contracts that do not maintain any state and that are deployed only once at a specific address. 
%In our study, 
52\% (15) of our participants (5 early-stage and 10 experienced developers) talked about using code from open source libraries such as OpenZepplin, a popular implementation of the ERC-20 and other ERC standards. For instance, P24 said %mentioned using OpenZepplin as his individual best practice in case of code optimization. 
\textit{``%I tried to optimize my data structure and the loops because of the gas fees. I try to keep my contracts as simple as possible and as short as possible and so 
I try to use the libraries from OpenZepplin because they are kind of best practice, for example, for access control and things like that.%, I use OpenZepplin libraries a lot.
''}

\noindent \textbf{Experience in code review.}
%\tanusree{ since the code review task was clearly about security (since it was after the interview),did you find out if they had ever reviewed code for vulnerabilities? And report which ones had?} \update{done}
%\yang{distinguish code review in solidity and in other languages. first talk about code review in solidity and then in other languaes.} \tanusree{done}
76\% of our participants had different levels of Solidity code review experience. Some conducted code review by pen and paper (P1,P6,P20, P25), some used automated test scripts/tools (e.g., OpenZepplin ERC20 library) for code review (P2, P24, P26, P27), and some used both techniques (P14, P19, P24, P26). %For instance, P24 stated, \textit{``In code review I prefer to find all the functionality and then focus on these things. So sometimes I go and check if there is risk for Reentrancy attack, if I've added for example the OpenZepplin library to control that and look for major risks.''}
Seven out of 10 early-stage developers had prior code review experience in Solidity. P1, P3 and P7 indicated that they had experience but are not experts. Four out of 15 experienced developers did not have prior code review experience in Solidity. However, among the seven participants who did not have code review experience in Solidity, three of them had much code review experience in other languages, such as Python (P3), RUST (P8) and Golang (P22). Four participants (P4, P5, P13, P17) did not have any prior code review experience in any language.

%\yang{which participant(s) had no code review experience in any languages?}

\subsection{Perceptions of Smart Contract Security}

\noindent \textbf{Is security a priority in their smart contract development?} %\tanusree{4.1.3 Given the prevalence of auditing in your findings, I would have liked to have understood more about the practice. Is it tools based, contracted out to experts, both? for example. "Security is a main objective for few developers" - I don't think this is the right framing given the largely qualitative nature of this study and it's small population size. It would be more accurately stated as "Security is a main objective for a few developers".}
%Different security adopters may consider security at varying degrees. 
%\yang{are they all experienced developers?} \tanusree{3 out of 4 are experienced. updated}
% 
We asked our participants their priorities in smart contract development, most participants (83\%) did not claim security as a top priority and cited three main reasons: (1) they need to ship projects fast and security becomes secondary, (2) their projects forked other popular projects (e.g., Uniswap), which are often already vetted by the community, and (3) someone else internally or externally will conduct security audits. However, the majority of participants considered functional correctness (66\%) and gas optimization (48\%) as important priorities. Both are actually closely relevant to security. We saw functional correctness as a form of security earlier. Gas optimization saves transaction cost but also reduces the possibilities of economic security issues such as  front-running (e.g., attackers observe a user's buy transaction in the mempool and submit another buy transaction with a higher gas fee to outbid the user's transaction). 

One early stage and 11 experienced developer participants mentioned security as one of the factors they considered in smart contract development. Figure~\ref{fig:pic-factor-comp} in the Appendix presents the different factors our participants mentioned. However, only five participants (P3, P11, P14, P26, P27) said that they considered security as a priority during development. Except P3, all other four participants were experienced developers. 
%\yang{Figure 6 it looks like 10+ experienced developers said security was a priority and 0 early stage developers did, is that right?} \tanusree{updated} 
They evaluated different aspects of security. For instance, P11 explained: ``\textit{I considered security the most important factor because you need to know the address in the blockchain to send and receive, [...] the address should be unique and it's secure and cannot be changed by someone else.%Then you need to consider the complexity because it affects the speed of transaction and speed is very important. So for me it's security and speed and gas optimization.
}'' %For ensuring security, these participants indicated manual inspection of execution traces in smart contract interaction flow. Additionally, Remix security plugins were used to identify known vulnerabilities and avoid obvious security risks. %This participant mentioned the main objectives for him is to verify the code's conformity to secure interaction followed by the complexity of gas optimization and speed.
% \yang{a general comment is that before or after you present a quote, you should elaborate it for the readers. for instance, here you can say the participant considers a few things important payment security, speed and gas fee}.  \tanusree{added}
%Similarly, P3, P14, P26 emphasized ensuring security from the very beginning, considering possibilities that the contract could be reentered or front-run by attackers at some point. 
P14 instead paid more attention to the reentrancy and front-running issues, and explained 
``\textit{%You know you've sort of created your system and now you're really putting a frame around it. And then from there going into a lot of gas optimizations to check if functions can be written more cheaply and going into
[...] a lot of security thought, can someone front run this? Can they reenter it? 
%You have to take a lot of consideration into the security and all the while also you would have written a lot of tests as well to check that your code is safe and it behaves as you expected.
}''  %Three of these participants were experienced developers.

\noindent \textbf{Why (Solidity) smart contract security is hard?}
%\yang{potential cut this subsection} 
Five participants (P25, P14, P18, P19, P24) pointed out that the Solidity language design has some inherent limitations for maintaining security. P25 (as well as P14, P18, P19, P24) noted %, \textit{``It's a poorly designed language. I'm hoping that there will be more programming languages that are safer and easier. The nature of program increases the attack surface.''} %\yang{why is solidity poorly designed for security?} \tanusree{participants mentioned about ambiguity. program doesn't differentiate state and virtual machine } \yang{we can discuss this in th chat}
%This participant further explained 
the ambiguity of Solidity programming language where function definition is not explicit: \textit{``It is very limited and so this requires you to go through hoops to do very basic things like access that you could do easily in [a] mainstream language''}. He suggested that adopting functions like in Python. P19 spoke about another issue with Solidity: \textit{``Contract work[s] like state machine when you send a transaction. It only appears like state changes. But in regular program, you can differentiate read only calls and state changes. Solidity can not do that.''} He also added the difficulties of performing proper string and array manipulations due to a lack of direct language/library support in Solidity. To fill this gap, this participant created his own assembler to correspond between Solidity instructions and machine code instructions. 

%\yang{1. participants mentioned about ambiguity. solidity program doesn't differentiate state and virtual machine. And EVM actually mixed of vm and state machine; 2. this language is a composition of containers of different languages which makes it difficult to handle security issues}

\subsection{Practices of Smart Contract Security}
Our participants explained their practices of ensuring smart contract security. They shared their methods and their experiences in working with other team members. We start by discussing the different strategies they followed. %In addition, they demonstrated their first-hand experience with different security methods and their rationale of making different security decision throughout the development life cycle.

%\yang{cut the following subsections on perceptions.}

\begin{figure*}[!t]
\centering
\begin{minipage}[b]{0.45\linewidth}
 \includegraphics[width=0.8\columnwidth]{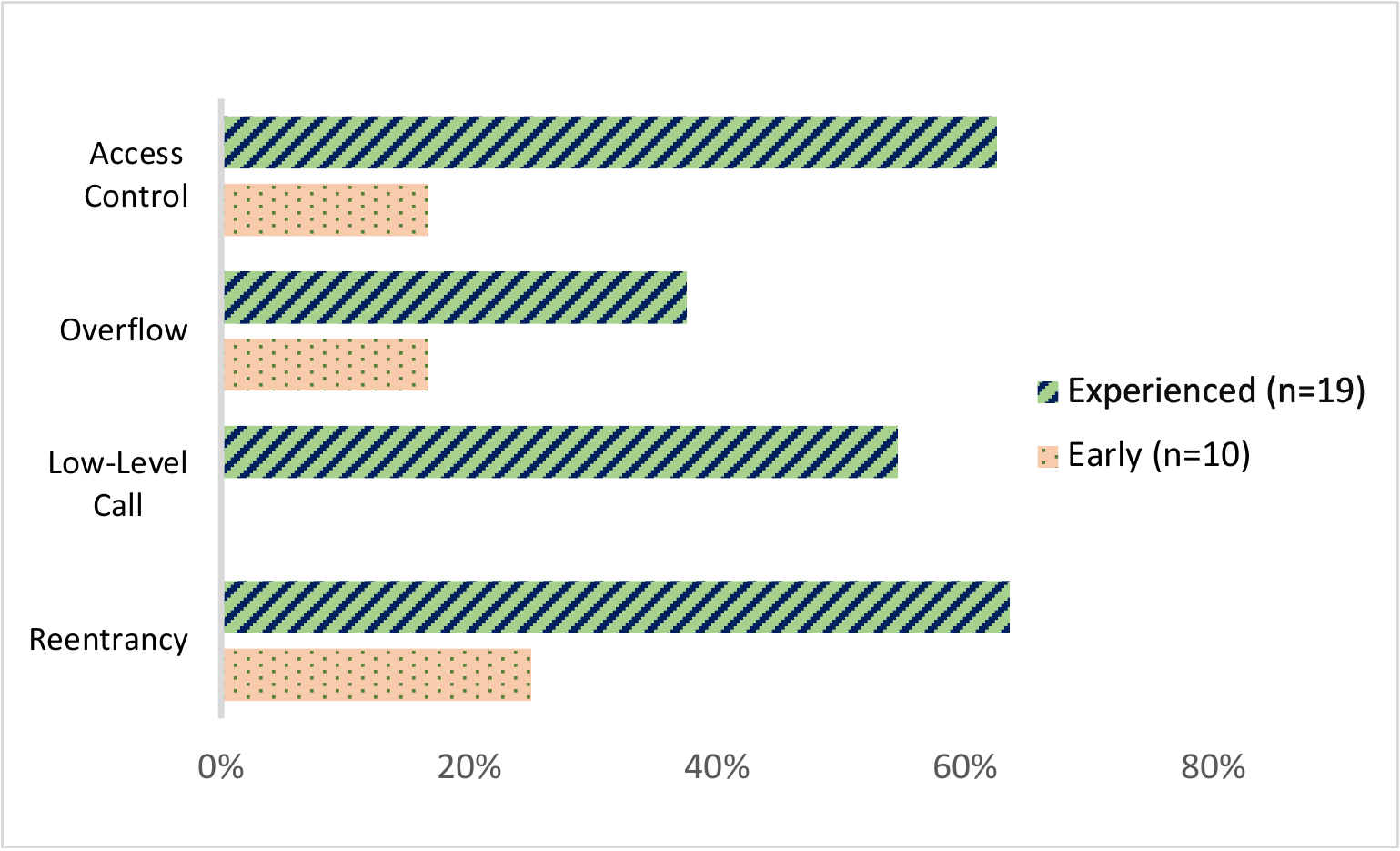}
\caption{Participant success rates of identifying security issues. The average success rates for early stage and experienced developers are 15\% and 55\%, respectively. %\yang{can you label the legend with the sample size N for early and experienced? can you also put fig 1 and 2 side by side, across 2 columns, either at the bottom of this page or the top of the next page?}\tanusree{done} 
%\yang{great, can you add pattern to the two colors. it might be hard to tell in grayscale}
}
   \label{fig:fig-s}
\end{minipage}
\quad
\begin{minipage}[b]{0.45\linewidth}
 \includegraphics[width=0.75\columnwidth]{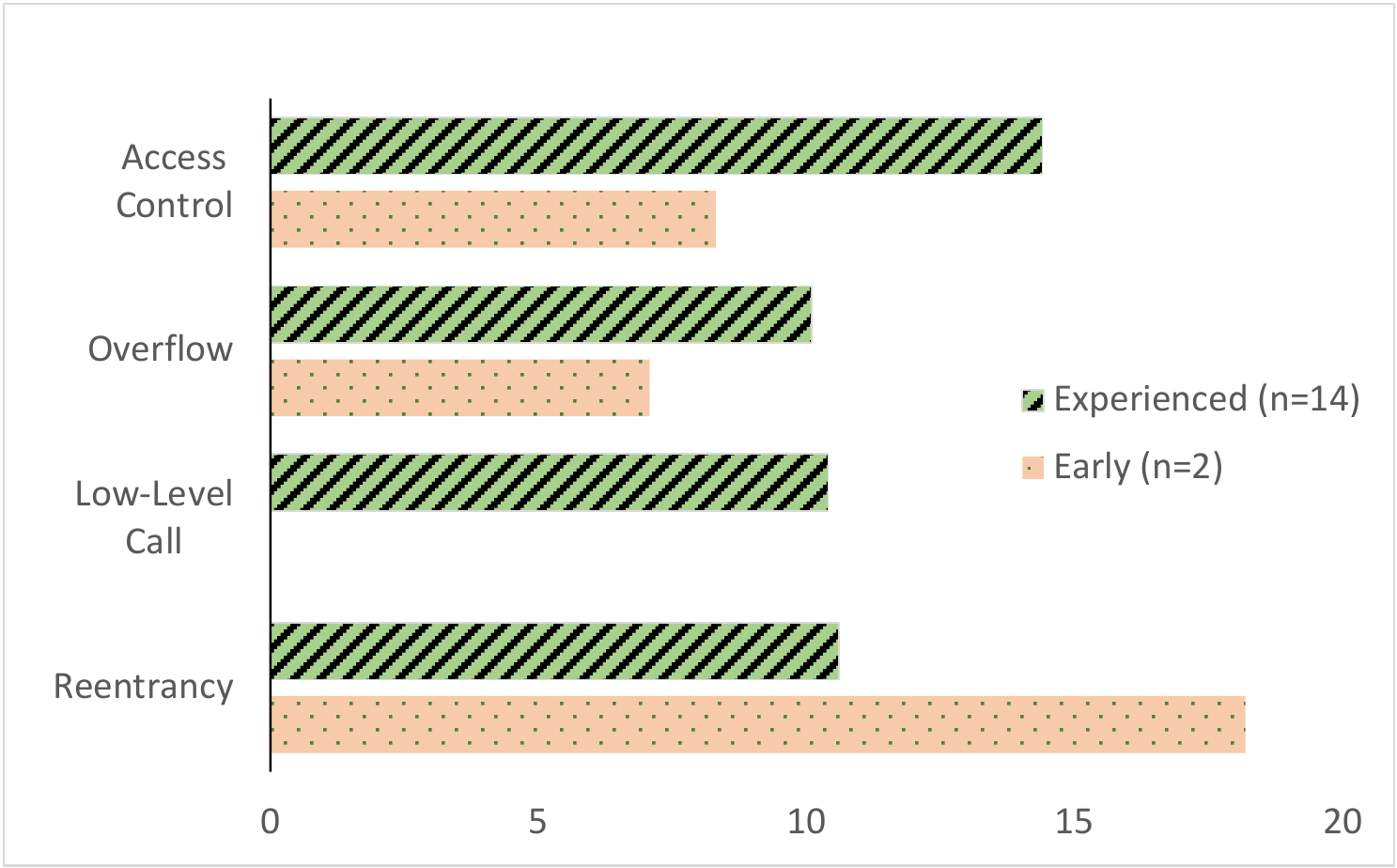}
  \caption{Participant average time (minutes) to successfully identify security vulnerabilities. %There were two early stage and 14 experienced developers who identified at least one vulnerability. 
  The early stage developers did not identify the low-level call vulnerability. 
  %Only one early stage developer was able to identify each of the three vulnerabilities.
  %\yang{the patterns look nice, but we need to increase the font size legend and axis labels for fig 1 and 2.}
  }
   \label{fig:fig-s1}
\end{minipage}
\end{figure*}

\noindent \textbf{Manual inspection of smart contracts}. 
%When our participants did consider and work on smart contract security, they used a variety of strategies to ensure security. Some leveraged tools such as static and dynamic analyses, while others mainly focused on manual code inspection. 
% 
%\yang{how many? Eight participants.}\tanusree{yes} 
Eight (28\%) participants reported using manual code inspection as the main method for identifying security issues in their smart contracts. 
For instance, P28 tried to think as an attacker while he read  through the code. \textit{``%It was mostly just trying to think of an adversary. 
Reading through the code and then thinking of how to steal everyone's money or do something malicious.''} 
As an other example, P6 combined flow charts with manual code inspection: ``\textit{I think step one always is to draw flow chart of where information are flowing and where [...] coins can get stuck and protocols. To find places where reentrancy can happen. So, having a graph of good representation can solve a lot of logical issues. And then trying to look at problems how calls are being made? Is there contracts gonna run out of gas? All that is just manual code inspection. %If graphical representation is precise or fine grained enough, [...], its the only way so far to do a security review correctly.
}'' As P6 read through the code, he created flow charts to help track the system logic but also identify issues of reentrancy and gas cost. %This approach conveys the importance of having a flow diagram of the contracts to be well aware of building blocks for identifying possible attacks scenario. 

\noindent \textbf{Use of smart contract testing or security tools}. %\yang{move this sub-section to "Practices of Smart Contract Security" section}
%\yang{how many? participants} 
Nine (31\%) participants reported using smart contract tools for security purposes. Testing or security tools, 
%In terms of smart contract security assessment, participants mentioned a small number of tools (both early-stage and experienced developers), 
such as static analysis tools (17\%), Truffle testing suite (14\%), Remix security plugins (14\%), MythX (7\%) and Slither (3\%). %Many (P1, P6, P16, P17, P20, P22, P25, P28) participants mentioned using pen-and-paper, white board or manual inspection besides the existing tools (
% Table~\ref{Tab:tab-new} in the Appendix shows their methods for checking security issues. 
%They felt the existing tools fell short for edge cases and were hard to use.  
%~\ref{code_review}). Most participants did not know about this kind of tools.

%\yang{these results can lead to many tool design implications and future work}
\noindent \textbf{Limitations of current smart contract security tools}.
%Several participants felt the tools fall short and they often just rely on manual inspection of the code. 
About one third of our participants (P4, P6, P7, P12, P14, P19, P22, P24, P25) spoke about the limitations of existing security tools for smart contracts.  %\yang{discuss these issues in design implications, e.g., how to make these testing tools more usable. this is a direction for future work.} 
For example, P6 explained: ``\textit{I don't think there are any automated tools [...] %majority of them come down to just manual inspection of the code and trying to run through possible execution traces. 
Even once you deploy the contract, having to write a constructor, then testing that contract is a really cumbersome activity. It's non-trivial, and there really exists no tool to help. Even truffle testing is quite hard to use in my experience.}'' 
P25 also complained about the Truffle tool being slow: \textit{``I didn't like truffle because it was a little slow. I had to rely on global imports into Files and then had to run all files from command line using truffle suit for testing which make iterative development complex.''} P24 similarly stressed the importance of better integration of security tools into the development process, saying ``\textit{I don't like to write code in an IDE and use another tool to test. I like my compiler to do everything for me. I don't like to use extra tool for checking bugs and security.}'' Better understanding and improving the usability of smart contract security tools is a direction for future research. 

After presenting how our participants self-reported their security practices, next we will cover how they actually behaved in identifying  security issues in the code review task.

\subsection{Findings from The Smart Contract Code Review Task}
%\textbf{RQ3: What is the level of expertise of developers to identify and solve security problems in smart contract code?}

%Now, we present the results of how our participants performed in the smart contract code review task. 
% Participants were generally successful in identifying and suggesting improvement for intended functional errors while their security vulnerability  identification results varied. %Study participants experienced very different rates of task success rate in identifying security vulnerabilities in smart contracts that they were assigned.
%Next, we present how our participants actually tried to identify security vulnerabilities in the code review task. 
In this task, we asked our participants to review a smart contract to identify any security issues or areas of improvement. We also asked their rationale for any code  modifications.

\noindent \textbf{Task performance of identifying security vulnerabilities}. 
%\tanusree{4.2.1 - perhaps personal reviewer preference, but given the small numbers, I would have liked the "n=" along with the percentages in the first paragraph."Only 21\% of participants...." - same comment as "few". This isn't quantitative; remove the "Only".} \update{done}
%We discuss the extent to which participants were able to identify security vulnerabilities during the code review tasks. We observed a wide variance in performance in the vulnerabilities identification and areas of improvement results across different participants. The category of outcomes includes, a) correct identification, b)failed to identify, c) identified false positive vulnerabilities. Besides reporting the success rate, we explain people's process and tool use during code review; observed challenges/ reasons why they failed to identify vulnerabilities or incorrectly identify vulnerabilities. We also described our observation of any possible association of participants' self reported years of experience, security background and occupation with task performance.
Overall, 55\% of participants found at least one security issue in the smart contract. %13 participants failed to identify any vulnerabilities in the smart contract. 
28\% (N=8) of participants identified both (all) vulnerabilities in the smart contract. More specifically, Reentrancy had the highest identification success rate (53\%), followed by improper access control (43\%), unchecked low-level call (40\%) and lastly integer overflow (29\%). 
%\yang{draw a bar chart of these success rates, two bars for each bug, one experienced one unexperiened} \tanusree{added} 

Figure~\ref{fig:fig-s} shows our participants' performance (success rate) in identifying different security vulnerabilities. We observe that experienced developers had much higher success rate (55\%) than their early stage counterparts (15\%). %We can see that experienced developers had higher success rate (54.5\%) (\yang{list the average success rate for experienced developers here}) for identifying vulnerabilities than early stage developers (14.6\%). %\yang{list the average success rate for experienced developers}. %\yang{this is whether they identified the bug, what about whether they fixed the bug?} \tanusree{this is the percentage of identifying vulnerabilities}
Table~\ref{Tab:tab-per} summarizes each participant's task performance as well as the method they used to identify the security issues. %\yang{move this table to Appendix} \tanusree{placed in appendix}.

In addition, we recorded and calculated how long participants took in successfully identifying each security issue, as summarized in Table~\ref{Tab:tab-time}. We observed that for reentrancy early stage developers spent more time. None of the early stage developers identified the low-level call vulnerability. Only one early stage developer (P28) %\yang{P29 has 3-5 years of experiences according to the participant table? double check} \tanusree{corrected, its P28} 
identified the overflow (about 7 minutes) and insecure access control issues (about 8 minutes). Code Review task time is detailed in Table \ref{Tab:tab-time}, Appendix \ref{code_review}. 
For participants who failed to identify any security issues, they all used up the 25-minute period we gave them. 

\noindent \textbf{Developers with formal training in smart contracts were associated with better task performance}.  
We also explored how our participants learned to code smart contracts and observed a trend that those participants who had formal educational training in smart contracts tended to do better in the code review task and had more awareness of security practice. We found six out of eight participants (who identified both security vulnerabilities in code review task) had formal training for writing smart contracts. Some of them ($P12, P25, P27$) designed courses on smart contract and has been teaching in universities and community groups. P27 said - \textit{``I learned smart contract coding as a part of my research in school. Then I helped design and develop the technical content to lecturer where I was the one to initiate that course in a college in Canada.''} 
%Similarly, P28 explained his smart contract coding experience through a formal class \textit{``I started looking into blockchain that led me into a theorem through a class, 3 years ago. From the theorem I learned about smart contracts, and started doing some research on hacking, and started writing smart contracts.''} 
%P12 is another participant who didn't have formal learning for coding, however, emphasize on security awareness in smart contract -\textit{``I am ambassador to central network at Ethereum Foundation and working as a blockchain Mentor. I'm doing activities to increase awareness of security in smart contract in Arabic group.''} These participants also shared a similar concept of threat model to mitigate risk while writing smart contract. P25 said- \textit{``security considerations go super slow in every single line of code. Think about how how things could go wrong, think about threat models and ways to to mitigate risk''} 
P28 mentioned formal learning for smart contract coding is important to avoid known vulnerabilities. \textit{``There's always new developers who are just throwing code out there as quickly and fastly with very little testing, no peer code review, no security considerations, and no experience and learning. So they're introducing problems. A lot of seasoned developers with awareness stopped doing this ages ago, and there's always new developers popping up, launching code with known vulnerabilities in them.''} We also found 10 out of 13 participants (who failed to identify any security vulnerabilities in code review task) relying on informal learning (e.g., YouTube, Stack Overflow, adhoc google search, Solidity docs) for smart contract coding.  %They seem to take courses for rapid skill development, and might overlook security aspects.
Future research can further investigate the relationship between formal training and security practices of smart contracts.

\noindent \textbf{Different approaches to code review}. Participants used different ways to identify security issues during the task. Most of them manually read through the code. Only four of them used security assessment plugins and/or static analysis tools. 

\begin{figure*}[!t]
\centering
\begin{minipage}[b]{0.45\linewidth}
\includegraphics[width=1.0\columnwidth]{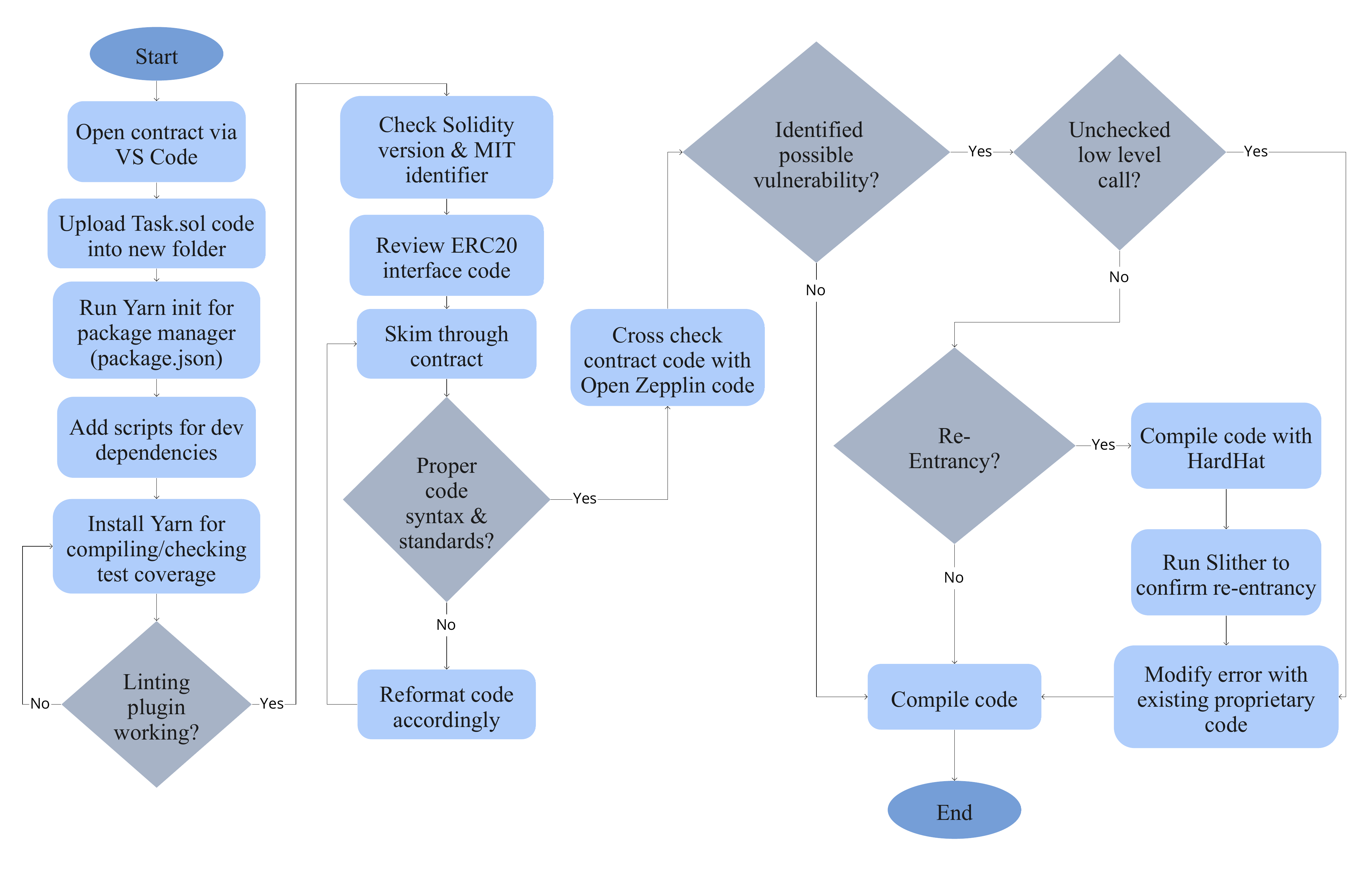}
\caption{P14's process for the code review task1. He used linting and static analysis tools.}
\label{fig:flow-dia-1}
\end{minipage}
\quad
\begin{minipage}[b]{0.45\linewidth}
\includegraphics[width=1.0\columnwidth]{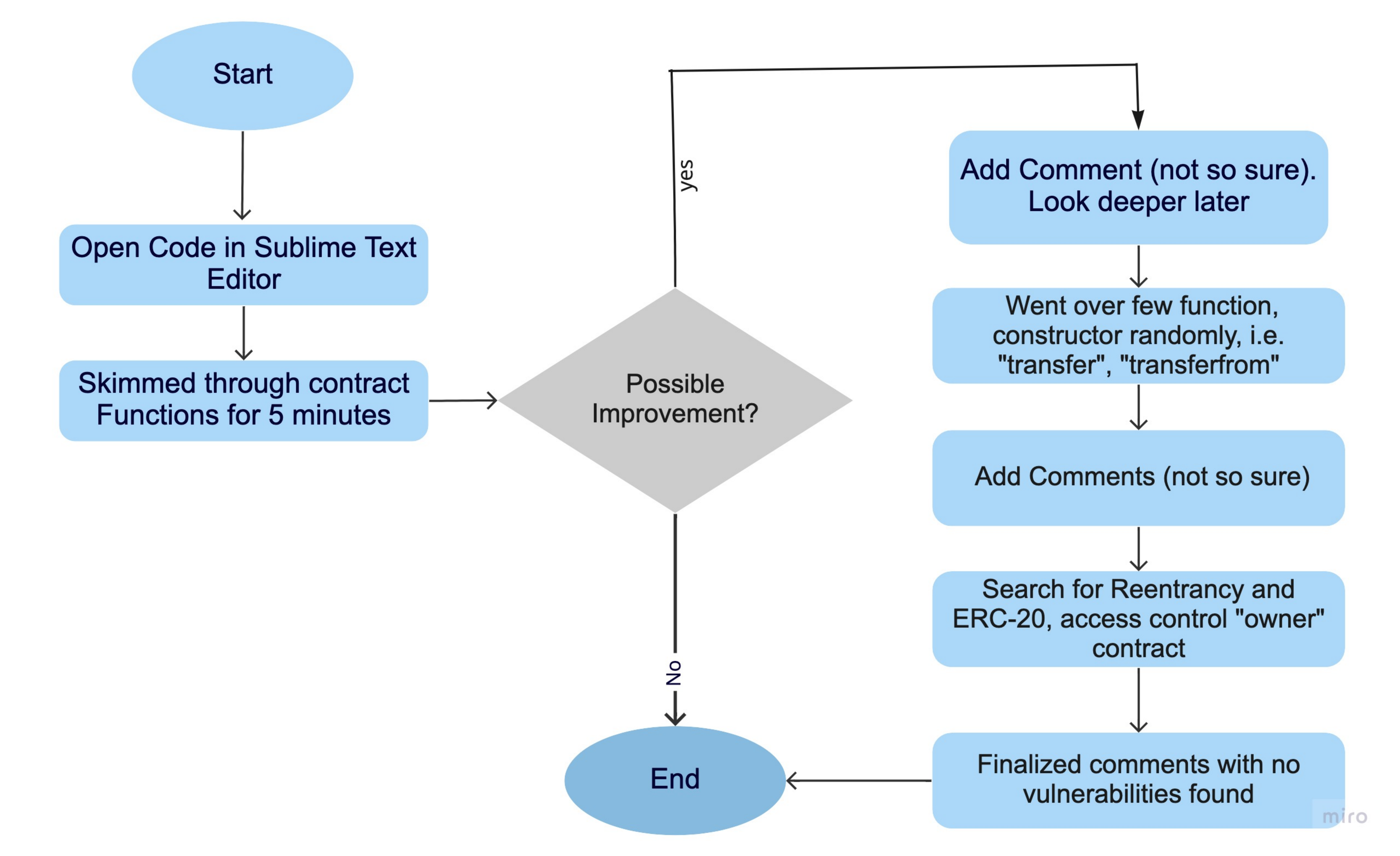}
\caption{P2's process for the code review task1. P2 failed to identify any security vulnerabilities.}
  \label{fig:flow-dia-p2}
\end{minipage}
\end{figure*}

For instance, P14 used linting and static analysis tools in his code review. He successfully identified both vulnerabilities (reentrancy and unchecked low-level call) in the contract. We did a hierarchical task analysis \cite{stanton2006hierarchical} %\yang{cite}
to break down how a participant conducted the code review into detailed steps. Figure~\ref{fig:flow-dia-1} %and Figure~\ref{fig:flow-dia-p2} 
illustrates P14's code review process.

% \begin{figure*}[!t]
% \centering
%   \includegraphics[width=1.0\columnwidth]{f1-14-final.pdf}
%   \caption{P14's process for the code review task, in which multiple tools were used for linting and static analysis.}
 
%   \label{fig:flow-dia-1}
% \end{figure*}

P14 started by quickly setting up his development environment, where he created a folder with a package manager and installed development dependencies so he could compile the contract. He stressed the importance of having the folder ready to compile first. He explained,``\textit{
%I'm gonna have a package manager quickly. This stuff is quite important. 
Let's get some scripts, dev dependencies and I'm going to run that here and run yarn. Things that will allow me to compile, check coverage, test, clean it. I will make sure that it's linting as well. 
%Once the whole code is compiling, I will start making improvements.
}''

After compilation, he started looking for possible security issues. He also manually went through the Open Zeppelin implementation of ERC20, ``\textit{I would literally track this whole thing and I'd use Open Zeppelin implementation. I can go through this implementation and see if it's good or not.}'' Importantly, he did not blindly take or trust the Open Zeppelin implementation even though it is widely used. After manually checking the code, he explained some of the possible vulnerabilities. For instance, he pointed out a potential reentrancy vulnerability in the withdraw function: ``\textit{it is withdrawing if the amount is less than the amount to just return false and subtracts the amount before it does the accounting before it's actually sending anything out, which is pretty crucial for preventing someone re-entering the function, which would be really bad.}'' He then ran static analysis with Slither to confirm his suspicion. He interpreted the Slither output: ``\textit{we can see if there's any fatal errors. Yeah, so exactly- state variables written after the re-entering pool. That's only if it's not successful. so I mean there's reentrancy in over here.%Whether it could be exploited, we'll have to figure out. So withdraw is probably the function you want to look at the most anyways.
}''

Unlike P14, many participants just did manual inspection of the code without using any security tools. %P6's approach of the code review task ($Task2.sol$) was quite different. %He preferred manual inspection of the code and think through possible execution traces. He did not use any security tools. %  the security plugins of the IDE (remix) that he was using to review the code. P6 also mentioned that ``\textit{I don't think there are any automated tools that, I think even once you deploy the contract, having to write like a contractor then testing that contract is a really cumbersome activity. I think it's nontrivial, and so there really exists no tool to help.}'' Wherefore instead of using any automated plugins or tools, 
For instance, P6 only did manual inspection. But he also searched information resources such as security best practices in the official Solidity documentation and the Open Zepplin documentation. He also successfully identified both vulnerabilities (integer overflow and improper access control) in the contract (task2) he reviewed. For instance, upon manual inspection, he found that 
%In ~Figure ~\ref{fig:code-snippet-2}, he stated that there had to be a safe math operation, because it could be overflow these in batchTransfer function. He indicated that
the amount of a local variable is calculated as the product of two other variables $cnt$ and $\_value$. %Therefore, having two $\_receivers$ passed into $batchTransfer()$, 
The later can have an extremely large value which can overflow the product and make it zero (as shown in Figure~\ref{fig:code-snippet-2}). 
He suggested using the Safe Math library. %, ``\textit{reuse the code that already works wherever you can. I think here I wanted to use the Safe Math library as well.}'' The Safe Math library is widely known and used to prevent math-related issues such as integer overflow. 

\begin{figure}[!b]
% \begin{minted}[breaklines,frame=single,fontsize=\small]{python}
% \begin{minted}{python}
\begin{lstlisting}[breaklines,frame=single]{python}
 function batchTransfer(address[] _receivers, uint256 _value) public 
 whenNotPaused returns (bool) {
    uint cnt = _receivers.length;
    require(cnt > 0 && cnt <= 20);
    uint256 amount = mul(uint256(cnt), _value);
    
    // below is possible taken care of by super.transfer 
    require(_value > 0 && balances[msg.sender] >= amount);

    // balances[msg.sender] = balances[msg.sender].sub(amount);
    for (uint i = 0; i < cnt; i++) {
        // checks if receivers are address(0x0) --> revert if true
        super.transfer( _receivers[i], _value);
        // balances[_receivers[i]] = balances[_receivers[i]].add(_value);
        // emit Transfer(msg.sender, _receivers[i], _value);}
    return true;
    }
}
\end{lstlisting}
\caption{Part of the contract code P6 reviewed and modified. The first few lines had a overflow vulnerability where local variable is calculated as the product of $cnt$ and $\_value$. By having two $\_receivers$ passed into $batchTransfer$, with that extremely large $\_value$, attacker can overflow amount and make it zero.}
\label{fig:code-snippet-2}
\end{figure}

P2, an early stage developer, was one of the 13 participants 
%\yang{earlier we said 55\% found at least one security issue, double check and ensure the results are consistent}\tanusree{corrected}  
who did not find any security vulnerabilities in the code review task. He was assigned the same task (task 1) as P14. Figure~\ref{fig:flow-dia-p2} illustrates P2's code review process. % performed containing reentrancy and low-level call vulnerabilities (Figure ~\ref{fig:flow-dia-1} and Figure ~\ref{fig:flow-dia-p2} are placed side by side to compare the code review flow of successful and failed attempt). % 
He started by skimming through different functions of the code. After almost five minutes, he said, \textit{``This seems like a security flaw here in ``transferfrom'' function. I will look deeper later.''} Then, he went through a few functions randomly, such as ``transferfrom'', ``allowance,'' and put another comment on the transfer function: \textit{``I did misread the ``transferform'' before. I was looking at in a wrong direction. Now it makes a lot more sense and then you're verified with a certain amount of tokens.''} At this point, he was not sure about vulnerabilities. For confirming some of his suspicion, he searched ``ERC-20 code'' and ``Reentrancy,'' in the Open Zeppelin contract example. He also read through some code examples in GitHub repos that included ERC-20 code example. As the time was close to 25 minutes, he concluded: \textit{``I don't see anything major. Not really seen doing much wrong.''} %See the flowchart of P2's code review task in Figure~\ref{fig:flow-dia-p2}. 

\begin{figure}
  \includegraphics[width=0.98\columnwidth]{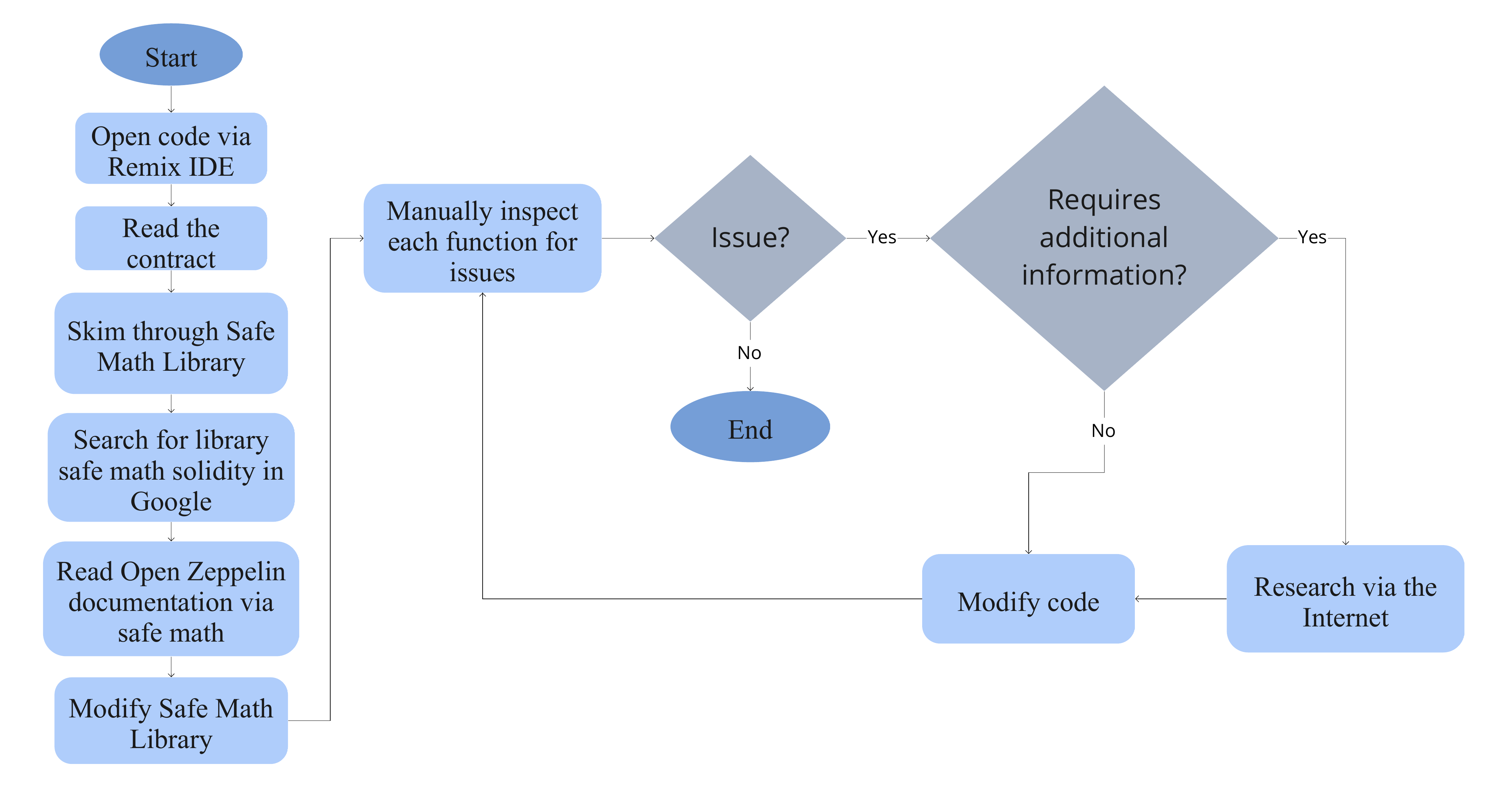}
  \caption{Code Review Task Flowchart for P6 who manually reviewed the code and successful in identifying both vulnerabilities in Task 2}
  \label{fig:flow-dia-2}
\end{figure}

What is interesting here is that even though P2 found and read some reference documentations about ERC20 (e.g., Open Zeppelin) similar to P14, P2 was still not able to identify the security issues. A similar example is P22 (whose review process flowchart is Figure\ref{fig:flow-dia-2} the Appendix). In addition, many of our participants (e.g., P3, P15, P20, P21) who did use security tool(s) also missed some security issues. These observations suggest that just by accessing standard documentation, reference implementations and security tools is not sufficient in helping developers (especially early stage developers) identify security issues.
We included a few more examples of how other participants performed the code review task in the Appendix~\ref{app:task-flow and time}.

\noindent \textbf{Code modifications for improvement}. %We also evaluated the modified code submitted by our participants. 
For the 16 participants who have successfully identified at least one vulnerability, seven of them modified the contract code for improvement and six (21\% of all participants) correctly fixed the vulnerabilities. %We checked their modified code to see if the modifications could possibly fix the embedded vulnerabilities. 
%\yang{how many successfully fix the problem?} \tanusree{done} 
The modified code snippets are included in the Appendix~\ref{app:Extra-code snippets}. Due to the time constraints, the other nine participants did not modify the code but verbally commented on how the contract code could be improved. Conceptually, all the comments would fix the security vulnerabilities.%\yang{how many verbal comments would successfully fix the problem?} \tanusree{done}
% \begin{comment}
% \begin{figure}[!t]
% \begin{minted}[breaklines,frame=single,fontsize=\small]{python}
%  function batchTransfer(address[] _receivers, uint256 _value) public whenNotPaused returns (bool) {
%     uint cnt = _receivers.length;
%     require(cnt > 0 && cnt <= 20);
%     uint256 amount = mul(uint256(cnt), _value);
    
%     // below is possible taken care of by super.transfer 
%     require(_value > 0 && balances[msg.sender] >= amount);

%     // balances[msg.sender] = balances[msg.sender].sub(amount);
%     for (uint i = 0; i < cnt; i++) {
%         // checks if receivers are address(0x0) --> revert if true
%         super.transfer( _receivers[i], _value);
%         // balances[_receivers[i]] = balances[_receivers[i]].add(_value);
%         // emit Transfer(msg.sender, _receivers[i], _value);}
%     return true;
%     }
% }
% \end{minted}
% \caption{Part of the contract code P6 reviewed and modified. The first few lines had a overflow vulnerability where local variable is calculated as the product of $cnt$ and $\_value$. By having two $\_receivers$ passed into $batchTransfer$, with that extremely large $\_value$, attacker can overflow amount and make it zero.}
% \label{fig:code-snippet-2}
% \end{figure}
% \end{comment}

\noindent {\bf False positives}.  
We also observed that sometimes our participants incorrectly thought they identified security vulnerabilities, which were actually not vulnerabilities (i.e., false positives). Four (14\%) participants had false positives and they tended to have less prior experience with smart contract development. For instance, %P1 indicated a bug in \textit{approve} function. Though he didn't mentioned what is the exact bug in there- only suggested that allowed has to add on to how much ether is allowed and in his opinion, there is an unexpected behavior in that function. 
%\textit{``If I'm not wrong, I think there might be a bug here in “function approve”, something is like warning lights are going off. But I think from memory it's something to do with multiple approve transactions. Then a miner will reorder them''} \yang{is this the race condition attack?} \tanusree{he was referring to race condition, but there was not any}. During the task, he didn't provide a modified code solution, rather he put comments in the codebase for this false positive vulnerability. \textit{``//Might be a vulnerability. Something to do with miners reordering the approve txs. // I kinda remember the approved function having “set to 0” kind of logic.''}. Similarly 
% 
P1 thought there was a security issue in the ``approve'' function (Figure~\ref{fig:P1-false} in Appendix). %\yang{it'd be nice to show the code snippet in the appendix.} \tanusree{added} % \yang{would be nice to include the code snippet here} \tanusree{didn't modify code}. 
He explained, \textit{``it is possible for adversary to call the function in StandardToken and this would bypass `whennotpaused' modifier.''} However, there was a ``Pausable'' base contract, which can implement an emergency stop mechanism where ``whenNotPaused'' modifier was only callable when the contract is not paused. Also when it is called by the owner to pause, it triggers the stopped state. Hence, one cannot bypass the ``whennotpaused'' modifier. %\yang{briefly explain why this isn't a security bug} \tanusree{addded} 
P1 googled different keywords related to ERC20 and finally found the Open Zeppelin ERC20 contracts for reference. %Generally, P1 was not familiar with common resources of smart contract security. 

% \yang{we should include relevant code snippets of the tasks in the appendix  \tanusree{added}+ upload all the contract code in an anonymous github account and put the link in the paper} 
%\yang{not absolutely required, but after we finish a full draft, we might need to measure/report the time that each participant worked on a task}

\noindent \textbf{Resources used during the task}. 
During the code review task, participants were told that they could use any resources or tools. Most frequently searched resources include Solidity documentation (28\%), Open Zepplin (17\%), and Google search with keywords (17\%). %Some of the other resources used were different GitHub repositories for ERC20, Medium blogs with specific queries (e.g. fallback function), $stackexchange.com$, and tutorials. 
Some of the search terms used were ``parent contract call,'' ``safe math parent class,''  ``fallback function,'' ``emit in solidity,'' %``Safemath Library code in openzeplin,''
``reentrancy vulnerability example,'' %``Sending Ether (transfer, send, call),''
``example of transfer and withdraw function,'' ``constructor example,'' and ``solidity version specific details.''

\noindent \textbf{Tools/Frameworks used in the task}.  
%\tanusree{regarding the vulnerabilities in the smart contracts of the review is if they are detectable by static analysis tools? I found it odd that only a very small number of the participants employed static analysis tools. Did you check if the tools are indeed difficult to be installed and used? Or should the developers only have to leave their IDE and run the static analysis tools from a command line to detect the vulnerabilities? Please also mention the reasons why none of the participants executed dynamic analysis tools.}\update{added explanation.}
%
After sharing the GitHub link of our smart contracts for the code review task, 
participants were initially using Remix (48\%), VS code (28\%), Sublime Text Editor (14\%) and IntelliJ IDEA (3\%), or manually conducting code review on GitHub (7\%) for the task.
% \yang{let's also provide the stats on the percentage of participants who used any static or dynamic analysis tools} \tanusree{added} 
%Based on their regular practices, they then added some plugins into their development environment for formatting code, creating graphs of contract interaction, and finding known vulnerabilities. 
In addition, 24\% of participants used a static analysis tool, such as Remix static analyzer plugin, Slither, and Oyente, during the task. One participant used a dynamic analysis tool, MythX %\yang{examples?}\tanusree{added} 
during the code review task. %, which might be because of the limited time they had. 

\section{Discussion}
\label{sec:discussion}

\subsection{Recap of Major Findings}

Our study aimed to understand the security practices and challenges of smart contract developers. Overall, we found that our participants had diverse perceptions and behaviors regarding smart contract security. 

%\noindent {\bf Early stage vs. experienced smart contract developers.} 
%We had 10 early stage (less than one year of experience) smart contract developer participants and 19 experienced developer (2-5 years of experience) participants. In general, the experienced developers tended to have more awareness of smart contract security (e.g., vulnerabilities, best practices, resources, tools) and perform better in the code review task than their early stage counterparts. 

While the majority (83\%) of all participants did not mention security as a priority in their smart contract development, a large portion of all participants mentioned functional correctness (66\%) and gas optimization (48\%) as important priorities, both of which are closely relevant to smart contract security. 

%Our early stage participants tended not to treat security as a priority in their smart contract development. 
%11 experienced and one early stage developers indicated security as one of their considered factors, however, not the top priority. 
Five participants said that they considered security as a priority in their smart contract development. Among the five participants, only one of them was an early stage smart contract developer while the other four were experienced developers. 
%\yang{Figure 6 it looks like 10+ experienced developers said security was a priority and 0 early stage developers did, is that right? In the recap of results in section VI it says "One early stage developer named smart contract security as a priority, while four experienced developers held the same view"}\tanusree{updated} 
Early stage developers tended to think security is not important for their projects since they felt their projects have little impact in part because they are not deployed in the mainnet and thus are not an interesting target for attackers. %They might be the conception of the importance of security from risk perspective which was negligible for certain project. Besides, 
%This kind of perceptions might result from a lack of security education and awareness of relevant information sources for beginners. 
% 
As a result, they often lack awareness of smart contract security issues as well as resources and tools about smart contract security. While the impact of security vulnerabilities in smart contracts deployed on the testnet is limited, there is a missed opportunity to motivate and educate (early stage) developers about smart contract vulnerabilities and corresponding best practices. Prior research in traditional software development suggests that this kind of perception can have a long-term negative effect on development practice~\cite{assal2019think}. Prior work on software security has also shown the value of improving security education and awareness~ \cite{acar2017comparing,gorski2018developers}. %``Security by design'' can be applied to smart contract development.}

In comparison, our experienced developer participants indicated that their companies or projects have faced security issues, including known and edge cases. For security practices, they mentioned using best practices and resources, such as Open Zeppelin contracts (reference implementations) and Solidity official documentation. However, their responses also suggested that they tended to rely on security audits as the final security evaluation.

%Participants in our interviews mentioned a number of development tools (e.g., Truffle, Hardhat, Waffle), and their practice of code reuse from open source projects and libraries. Early stage developers emphasized on code-reuse for convenience. More than two thirds of our participants had different levels of prior code review experience. 

In terms of their actual behavior in the code review task, early stage developer participants had a much lower success rate (15\%) of identifying security vulnerabilities in the reviewed code than their experienced counterparts (55\%).  %\yang{why? is it because early stage developers don't know the security vulnerabilities or they have heard of them but don't know how to identify them} 
The majority (72\%) of all participants only did manual inspection of the code where they tried to run through possible execution traces to find out vulnerabilities as well as edge cases. 
20\% of early stage developers used security tool(s) in the code review, compared with a higher percentage (32\%) of their experienced counterparts. The early stage developers only used Remix with its security plugin whereas the experienced developers used more tools such as static and dynamic analysis tools (e.g., Slither, MythX). 
% 
%We also observed a wide range of practices of how smart contract developers reviewed smart contract code and identified security vulnerabilities. 
%Some participants from the DeFi industry and with a research background mentioned that security tools were not efficient and usable enough, and thus they preferred to manually inspect smart contract code. 
% 
%Few participants used static analysis tools such as Slither during the code review to only point out known vulnerabilities. 
Our hierarchical task analysis of their code reviews implies that just by accessing standard documentation, reference implementations and security tools is not sufficient. Many developers checked those materials or used a security tool but still failed to identify the security issues.

In addition, one third of the participants pointed out many limitations of smart contract security tools, such as lack of comprehensive libraries, integration of these tools as part of the IDE as well as complex user interfaces. How to make  existing smart contract security tools more usable is an important direction for future work.

%Our study indicated that project leaders who specified system requirements lacked knowledge of smart contracts security. Due to lack of first hand development experience, their initial system requirements often ignored security and code optimization until that impacted their users. As one participant pointed out, \textit{``if they were more of a user themselves, they might have more empathy for gas costs and security.''} If these leaders are more aware of security-related implications, security might be more prioritized in these projects. \tanusree{maybe we can cut this paragraph since its not representative now for current result section}

%In those cases, developers who wrote smart contract in the first place suggested that education and experience could make a difference in smart contract development environment. One of the participants pointed 

\subsection{Smart Contract Security vs Traditional Software Security}

% \tanusree{
% % "For traditional software security, there are systematic software security guidelines. " - I don't see any reason that traditional security development lifecycles, threat modeling, or security/vulnerability/pen testing can all be applied here to.
% "Besides, the additional economic considerations (e.g., incentive mechanisms, gas cost) add more complexities to ensure smart contract security." - I didn't follow this. Examples would have helped support this statement. "can create new functionalities by composing different smart contracts" - did any of your participants mention this? I don't recall you calling out the composability challenges. -D}
%As discussed, some of our study results seem to resemble prior work's findings on general software security.
There exists systematic software security guidelines, NIST \footnote{https://csrc.nist.gov/Projects/ssdf} and OWASP \footnote{https://owasp.org/www-project-security-knowledge-framework/} to facilitate secure software development practices amongst business owners and software developers for traditional software. 
%Following these practices helps s
These help software producers to mitigate the potential impact of exploitation, and address the root causes of vulnerabilities to prevent future recurrences. Compared with traditional software, smart contract development is a relatively new landscape. With the recent boom of the DeFi industry (since 2020), a new pool of developers joined this industry and started writing smart contracts. Their knowledge of and experience with smart contract security vary widely, as observed in our study. 
%Recent research proposed a smart contract risk classification based on the NIST Bug Framework for the formal classification of smart contract vulnerabilities \cite{dingman2019defects}. %However, smart contract vulnerabilities are varied from traditional software security framework, which needs further research. Though 
There are some information resources on smart contract security, such as ConsenSys and Solidity documentations, which provide design patterns that developers should follow. However, we are not aware of a framework for organizations to properly assess and communicate smart contract security during the development life cycle. %Moreover, it is also challenging for new developers to learn the impact of security exploitation and mitigation strategies of vulnerabilities. 

%Throughout our research of the issues around lack of security guidelines for Solidity smart contracts, we have concluded that the security of smart contract in Ethereum blockchain  is volatile. It requires significant improvements to make it a reliable medium for developers to constructively learn  practices of developing secure smart contracts and taking risk mitigation strategies. 

\noindent \textbf{Apply software security solutions in smart contracts}.
Our results suggest some similarity of practices and challenges between smart contract and general software security.  Thus some security strategies of software development can be applied to smart contracts. For instance, 
%Some of the themes in smart contract security practices align with traditional software security studies. 
some participants indicated the importance of relevant documentations within the IDEs which can support their learning in smart contract development. The literature on secure software engineering has made similar suggestions (e.g.,~\cite{naiakshina2017developers,acar2017comparing}).  %stressed the significance of standards and its integration into popular frameworks so that it can help developers ensuring secure password storage. \cite{acar2017comparing} advocated the importance of proper documentation and comprehensive features covering a broad range of use cases. Integration of accessible and clear security documentation into go-to frameworks used by developers can also be adopted into the smart contract domain, which can potentially serve as security education resources. 
Our participants also desired well-structured, context-specific security warnings. This aligns with prior literature that suggests effective security warning to reduce cryptographic API misuse \cite{gorski2018developers}. 
%investigated security warning as the potential mechanism to reduce the negative implications of cryptographic API misuse. Integrated security advice during writing code is proved to have a positive effect on code security. This study also indicated that generating effective security advice as warnings is challenging and context-sensitive. Context-specific security suggestions and warnings can immensely support building a mature smart contract developer community. Participants in our study indicated different information sources and practices of re-using code. These approaches are handy, but comes with the possibility of vulnerabilities. For example, reusing code from sources like Stack Overflow can lead to vulnerabilities due to embedded insecure functions. Instead of randomly searching information, security warning can provide constructive suggestions of secure coding examples. This could support beginner-level smart contract developers to polish up their security perceptions and practices.
The prior literature on cryptographic APIs \cite{acar2017comparing} suggests that simplified libraries can promote security. However, it might not be the case in the domain of smart contracts. Many participants in our study explicitly requested the security/testing libraries to be more comprehensive in covering most if not all security issues. Further research is needed to investigate this idea.  %hypothesis about library complexity and smart contract security. %They also pinned down the gap in recognition of security problems by developers due to complex libraries. 
%Some participants in our study mentioned 
Our study also highlights the inconvenience of using different tools and plugins to detect different vulnerabilities. 

\noindent {\bf Why is smart contract security different and difficult?} 
Smart contract security have some fairly unique or more prominent characteristics and often have a significant financial impact, especially for DeFi projects. It is fairly easy to monetize the attacks (e.g., ``stealing'' tokens and  passing them to a mixer before selling them). In addition, the system states are visible on public blockchains and the smart contract code is often open-sourced, thus the public nature makes them targets for attacks. Besides, the additional economic considerations (e.g., incentive mechanisms, gas cost) add more complexities to smart contract security (i.e., economic security)~\cite{parizi_smart_2018}. %Perhaps because of all these factors, we have seen regular reports in the news about real-world smart contract projects being attacked. 
%Smart contract are implemented in distributed system and generally open in nature. This makes it accessible by public and increases the possibility of attacks. 
Furthermore, in the smart contract development ecosystem, there is wide reuse of library code or interactions with other smart contracts (DeFi projects known as ``money lego'' because they can be created by composing smart contracts from different projects) \cite{chen2021understanding}. This composability, however, also opens the door for trusting and calling untested and potentially vulnerable or even malicious smart contracts. 

%In our study, new developers often talked about the difficulty of starting contract development from scratch. Rather, they mentioned re-engineering some logics from existing projects for improved efficiency. Another frequent aspect discussed during study was formal verification. As per participants' responses, 
%In addition to mention of static analysis, some participants also felt that  many vulnerabilities could have been avoided with the help of formal analysis and verification, while Solidity was not built for formal verification.  %Some alternative modeling-driven techniques have been suggested for programming and formal verification centric solutions.  
%Future research and development could create more formal verification tool support for smart contracts. 

% ease of monetization of attacks, and the readiness to throw large investments  in these, are the main reasons security should be much more of a concern. 

\subsection{Implications for Smart Contract Security}
%\tanusree{ "A checklist of common smart contract security issues inside IDEs would be useful." - this is not phrased as a suggestion from participants, which is how you started the section. Is this really from them, or from you, based on things they said?
%"Two participants mentioned that they would like to have a feature that could show the storage slots while writing code" - from the explanation, I don't see how that's related to security.}
% most focuses on functionalities. don't pay enough attention on security and felt it's someone else's job. didn't understand the intricate relationships between code optimization, gas cost, and security. many vulnerabilities are general software vulnerabilities not unique to smart contract. the ones that are particularly salient in smart contracts are: reentrancy. 

% new developers unaware of useful resources. 

% \yang{anything?}
Our study results provide a number of implications for smart contract security education and tools. %, which can be  summarized into three ideas: a) better documentation, b) formal verification, and c) UI design and visibility of usable plugins.
Specifically, we will discuss the implications for smart contract education, compilers, libraries, development frameworks, IDEs, and testnets. 

%\subsubsection{Functionalities of Security Tools}

%\noindent {\bf Languages}.
%Our study focused on Solidity because it is the most popular smart contract language. However, some of our participants pointed out design limitations of the Solidity language (e.g., implicit function definition) that can present hurdles in writing secure smart contract code. Solidity has being actively evolved in part to mitigate some of these security implications, but in turn developers often have to deal with code under different versions of Solidity, which can be confusing or overwhelming. Having clear documentation of the security implications of these language changes would be helpful especially for less experienced developers. 

%There are also alternative smart contract languages such as Vyper, which is designed with security as a key requirement. However, it is unlikely that developers will choose Vyper over Solidity, which is more popular, unless there are more tool support and reference open-source implementations of ERC standards (e.g., ERC-20) in Vyper. 

\noindent {\bf Education}.
Our results suggest that accessing documentations, reference implementations and security tools is not enough in helping developers identify smart contract security vulnerabilities. Education that improves their motivation, knowledge and awareness regarding smart contract security is crucial. While there are many reading materials for educational purposes, we believe what is missing is hands-on exercises or labs for smart contract security, similar to the SEED projects for computer security in general~\cite{Du_theseed}. Once these educational materials (e.g., hands-on labs) are created, they should be brought up when corresponding security issues are detected (i.e., teachable moments) in the various components of the smart contract ecosystem (e.g., compilers, security tools, IDEs, testnets).

\noindent {\bf Compilers}. 
%
%\textbf{Better integration of security concepts in the development process}. %\tanusree{can we call it "Better Explainability of security concepts"}   
Our results indicate that most of our early-stage developer participants were not familiar with the basic smart contract security concepts and common vulnerabilities. Some suggested integrating security analyses (e.g., static analysis) directly into the Solidity compiler, so that they can obtain the security assessment without any extra step.

\noindent {\bf Code libraries}.
%
%\textbf{Improving security tools.} 
%Participants in our study frequently mentioned lack of mature security analysis tools in smart contract domain. We noticed many participants manually inspecting during code review task. 
Some participants commented that current security libraries fall short of covering edge cases. %They desired comprehensive security libraries in existing security tools for out-of-box vulnerability identification. 
Most of the recent tools work on the byte code for identifying security vulnerabilities. The parser and symbolic execution engine work solely on the byte code, so if developers see a potential integer under/overflow, they will report it, but they do not know where it occurs in the source code. A direct mapping between security vulnerabilities and source code would be valuable. %\yang{was this mentioned by participants? if not, leave it out} \tanusree{yes. mentioned by 3 participants}
%For comprehensive security assessment, there needs a combination of symbolic execution of byte code, fuzzing and static analysis. This combination can start matching known issues and vulnerabilities on the AST level and then widen to IR to match vulnerabilities more precisely. 

%\textbf{Improvement of security testing library} 
\noindent {\bf Error / warning messages}. 
Many participants noted the error messages from current testing libraries can be overwhelming and hard to understand. % and safety check for data types. Even EVM in some cases does not provide support to display error messages for certain failures. 
For instance, P23 felt the error messages lack actionable insights. %noted \textit{``Testing libraries tend to have pretty bad error messaging and it becomes incredibly difficult to decipher a lot of errors that you get when you're running tests, but that's not specific to make the correction in code. [..] This is the problem with testing and deployment libraries that don't really provide insights for error messaging and debugging.''} 
The (security) error messages can be improved by including links to detailed explanations, known incidents, and how to correct the issues. Since many early-stage developers did not see the importance of security in smart contracts, connecting their own code (vulnerabilities) to past incidents of security vulnerabilities might improve their awareness of smart contract security. Some participants also suggested having better GUIs to present those information. For instance, P7 compared the Solidity tools to that of another language Python, which has a nice graphical user interface to show where exactly the problems are in the code and how significant the effect can be. Creating hierarchical GUIs where developers can see an outline of issues and can then zoom into specific issue or information would make the security assessment outputs less overwhelming. 

%We also noticed in literature that EVM even cannot throw out the exception in certain transaction failures \cite{atzei2017survey}, which can lead to reentrancy attacks if related issues remain unresolved during development. Thus, for secure smart contract development, we need well-tested standard libraries to better support error reporting and safety checks of data types. %{\bf lib}

\noindent \textbf{Formal verification}.   
In addition to static analysis, some participants also felt that many vulnerabilities could have been avoided with the help of formal analysis and verification, which were not built into Solidity.  %Some alternative modeling-driven techniques have been suggested for programming and formal verification centric solutions.  
They believed that formal verification can enhance code coverage in terms of correctness, because they are based on mathematical proofs. Future research and development could create more formal verification tools for smart contract security.

\noindent {\bf Integrated Development Environments (IDEs).} 
Our study suggested integrating security concepts into IDEs. A checklist of common smart contract security issues inside IDEs would be useful. The security tools (e.g., static analysis) can show which common security issues from the list are present in the code. %For instance, a required built-in Linter  extension within the IDE can also mitigate these issues. The extension could show where the problem is in the code, as well as links to style guides on how to write the code more appropriately. Furthermore, if there is presence of security vulnerabilities, it should explain why it is a crucial vulnerability and what might happen if not dealing with it. 
Many participants liked the convenience of the web-based Remix IDE, but wished its features can match those in the desktop-based IDE (e.g., Visual Studio Code) where more security plugins are available. Some of them also desired the security plugins can be more discoverable, either by enabling them by default or providing better search capability (e.g., creating a list of all security-related plugins).

%Besides, they also desired interactive web/online versions of IDEs with better testing and compiling capabilities. One of the participants indicated the limitation of Remix's interactivity and functionality, and called for improvement. %{\bf IDE}

%\textbf{Better visibility of useful plugins}. 
%\noindent {\bf IDE plugin visibility.} 
%During the code review task, it was quite common that when participants used static analyzers and security plugins, it took time for many of them to search the intended plugins in their IDEs. Two participants mentioned that they would like to have a feature that could show the storage slots while writing code, because this is one of the biggest issues they would consider when they need to do contract sizing and limit gas use. \yang{how is this related to security?} \tanusree{this point not necessarily about security, its for a design suggestion for development} So visualization of storage slots would be extremely useful to them. However, they also mentioned that perhaps there existed some obscure plugins that were hidden somewhere in the IDE. In that case, better visibility of useful plugins in IDE would be helpful during writing or reviewing smart contract code. Including, enabling and running security tools by default as part of the contract compilation process in IDEs might be even better. 

\noindent {\bf Development frameworks.} 
Some participants mentioned existing development frameworks to be heavy-weighted and difficult to learn and use the security functionalities. For instance, P6 noted Truffle, a popular smart contract development framework is ``unwieldy'' to learn and run security tests. One possibility is for these development frameworks to include a ``security mode'' where security analyses are automatically done as part of the compilation process. %As one of the participants (P6) said, \textit{``It's hard to use tooling around security issues because they necessarily require a whole test environment. Truffle tries to create a good framework for testing and how to write tests, but it's just unwieldy. Even the testing tools that exist, are far too hard to learn.''}. %{\bf framework}

\noindent {\bf Testnets.} 
Many of our early stage developer participants said they did not consider security as a priority in their smart contract development because they mostly deployed their projects in testnets where the (security) bugs have no real impact. While this sentiment might be understandable, this observation raises the question of whether a separate testnet could be designed specifically for smart contract security education and testing. For instance, the testnet is deployed with hands-on labs, which include smart contacts that purposefully embed existing smart contract vulnerabilities (imagine a vulnerable version of OpenZeppelin ERC20 contracts) and developers need to reuse or interact with these contracts. This testnet could help developers experience and learn how to mitigate these security vulnerabilities.

%\subsubsection{Usability of Security Tools}

%\textbf{Easy to use Frameworks and IDEs}

%\noindent {\bf Tool UIs.} 
%Some participants wished the security tools have better user interfaces. They compared the Solidity tools to that of another language Python, which has a nice graphical user interface to show where exactly the problems are in the code and how significant the effect can be. For instance, P7 preferred to see where in the code has the security problem. %One of the participants (P7) said,
%\textit{``I would like to see a better UI where you can see what is wrong in your code rather than what static tools give so much information that sometimes it's just very overwhelming. [..]So if Solidity were to develop some tools then they could have a nice graphical user interface and show the problems to user more effectively to follow.''} 

\subsection{Limitations and Future Research}

%\yang{add study limitations. First, we had a fairly small sample size but our participants had a diverse set of experiences and backgrounds. Similar to any qualitative studies, we cannot claim the generalizability of our results to the broader smart contract developers. However, our findings do present the real practices of this community albeit we do not know how common these findings are. Follow-up large scale surveys can further answer the frequency question.} \tanusree{updated with 2 limitations}

Our exploratory qualitative study has many limitations. First, we have a small sample size (29 participants), and thus we cannot claim our results can necessarily be generalized to the broader smart contract developer population. However, our participants were smart contract developers with diverse levels of experience and backgrounds, and our findings do present the real practices in this community. Future research can conduct large-scale follow-up surveys to help answer more quantitative questions such as the frequency of different types of smart contract security practices. 

Our sample included many students, which one may wonder if they can represent some of the smart contract development practices. We note that ``smart contract developers'' are not bound to those who have a formal title as a (full-time) smart contract developer. All of our participants had experience in developing smart contracts and thus met our criteria of smart contract developer. 
In addition, the emerging smart contract (and DeFi) space has developers from very diverse backgrounds and with different levels of experience (including students), not just full-time developers. Many students are indeed creators of DeFi projects. In popular DeFi hackathons such as ETHGlobal and Encode, most participants were students. We see our diverse sample as a strength since it fits with the diversity of the smart contract developer population.

Our study focused on Solidity, the most popular programming language for smart contracts. 
%In this study, we limited our research to only Solidity smart contracts. 
There are other smart contract languages such as Viper and RUST. Further studies are needed to smart contract security practices in other languages. In addition, we only tested four common smart contract vulnerabilities in our code review task. %There are many more smart contract vulnerabilities. 
Future research is needed to investigate other smart contract vulnerabilities such as front running, flash loan attacks, upgradable proxy and delegate calls which may require developers to have a good understanding of the economic aspects of DeFi projects (e.g., incentive mechanisms of Ethereum, token economics, how token prices interplay with automatic market makers in decentralized exchanges). 

%Interview results indicate that smart contract development is still in its infancy: there is no generally accepted way to secure smart contract code; the existing development and security tools are still evolving; development and runtime platforms (i.e., programming languages, virtual machines) still have a lot of limitations; online learning resources and community supports are limited for security education. We also found frequent practice of code re-use that can be both the center of maturity and source of vulnerabilities due to lack of security education. From the code review, we observed a positive relation between the success rate of identifying security vulnerabilities and participants' years of experience; as well as a difference between individual and team security practices. Based on the findings, our  work leaves room for future work in multiple directions.

Last but not least, our participants were asked about security practices in the interviews before doing the code review task. Therefore, the interviews could prime our participants to think more about security in the code review and did better than they would otherwise. We also noticed a positive association between years of experience and successfully identifying security vulnerabilities in the code review task, this hypothesis however needs to be tested in future large-scale studies.

%In future research, one may conduct a large-scale lab experiment with both known and unknown security vulnerabilities to investigate the impact of different expertise level and habituation of smart contract developers on task success rate.

%Unlike the traditional software industry, we have seen smart contract developers from diverse background. It is clear that ``smart contract developers'' are not bound to those who have a formal title as a (full-time) smart contract developer. In fact, the emerging smart contract/DeFi space has developers from very diverse backgrounds and with different levels of experience (including students), not just full-time developers. Many students are creators of DeFi projects. In popular DeFi hackathons such as ETHGlobal and Encode, most participants are students. We see our diverse sample as a strength since it fits with the diversity of the smart contract developer population. %Our findings offer an indication for the ecological validity of smart contract security developer studies with students, and these results can be promising for future studies.

\section{Conclusion}
\label{sec:conclusion}

To understand smart contract developers' security perceptions and practices, we conducted an exploratory qualitative study consisting of a semi-structured interview and a code review task. We found that many developers rely on others or external audits to ensure security of their projects. When they do assess smart contract security, they often do it manually and find a lack of security resources and tools. Given the recent rise of smart contract projects (e.g., decentralized finance) and their associated security attacks, providing better educational materials and tool support especially for novice developers is paramount for the healthy growth of this domain. 

%Through the study, we identified several barriers which hurdles more secure smart contract development: under-estimated importance of security perceived by individual developers and companies, a lack of security learning resources, unusable security tools, etc. Based on the findings, we emphasized the importance of smart contract security, and propose design implications for more comprehensive and usable security tools.
%Bibliography
\bibliographystyle{unsrt}  
\bibliography{references}  
%\newpage
\section{appendix}
%\section*{Appendix}

%\begin{comment}

\subsection{Pilot Study}
\label{appendix:pilot}
We conducted a pilot study with 4 smart contract developers to test our study design including interview questions and code review tasks. %and validate the feasibility of the study by assessing the inclusion and exclusion criteria of the participants 
%\cite{in2017introduction}. %This was the first step for planning and modification of the main study based on our study experience, pilot participants' feedback and suggestions. In this process, we assessed recruitment potentials, effectiveness of interview questions and code review tasks to increase the researchers' experience with the study methods. 
% 
%In our study, there are two parts: i) semi-structured interview to understand experiences and practices around smart contract development and ii) smart contract code review tasks to discern the level of expertise of developers to identify and solve security problems in smart contract. Our interview questions includes queries on smart contract development background, challenges and current security practice, standards, method used/followed by the participants. 
For the code review task, first we implemented five smart contracts, each including one common smart contract security vulnerability (e.g., reentrancy, under/overflow, access control). We present details of these smart contracts in Section~\ref{sec:task-design}.  

%During the first two pilots, we randomly provided three contracts to each participant for review. We estimated 30 minutes for the semi-structured interview and 30 minutes for the code review task; 5 minutes for discussing the area of improvement on the code review tasks and 5 minutes for the exit interview to get feedback from participants. %Therefore, we estimated  hours for the entire study. 
\noindent {\bf First round pilot.} In the first two pilot sessions, we had participants who were doctoral student researchers in the blockchain and smart contract security area. We conducted the interview and provided three contracts to each participant for review.
They felt the interview questions were good. %Our first part of study (interview) proceeded well. We got responses which were expected outcomes for articulated research questions. 
For second part of study (code review tasks), %we had the following protocol for code review.
%\begin{itemize}
%\item We asked the participants to share screen, so that we could observe how they were approaching to finish the code review tasks.
%\item We asked the participants to think aloud while they were reviewing the code, so that we could understand their approaches and mental models better.
%\item We asked the participants to note down areas of improvement for security if necessary.
%\item 
we gave the pilot participants 10 minutes for each task (30 minutes in total for all 3 tasks).
%\end{itemize}
%
%
They provided suggestions on how to improve the code review task to make it more realistic. For instance, they felt the time was too tight to review three contracts, and suggested having each participant review only one contract, which should be more comprehensive and realistic by having common functionalities. They mentioned that code review takes an effort, and might be distracting without a realistic contract .
%Pilot participant A mentioned that \textit{``it's a little bit different from what I do that review for other contracts. So I usually being very quiet like I'll just write it down instead of speaking it loud. Now that it's a little bit different.''}. Participant A also mentioned that \textit{``I think it's a little bit short to do 3 code reviews in such a short time. I think making it longer would be better, because doing code review takes an effort and then actually it's like suddenly you get a task and so you don't have a really  good contract, so it might be distracting.''} . Participant B also suggested that \textit{``Give only one task, put like 20 minutes on only one task and then that would be better than 30 minutes focusing on three tasks.''}. 
Therefore, we created two basic smart contracts based on ERC20 token standard, which have basic functionalities to transfer tokens as well as allow tokens to be approved so they can be spent by another on-chain party. ERC20 is the most common token standard on Ethereum and should be familiar to smart contract developers. The length of smart contract code varies significantly: some library/interface code can be 20 lines, while other more complicated contracts (e.g., Uniswap router code) can be several hundreds of lines. Practically, to fit into the time frame of our study, the code cannot be too long. Therefore, our 2 newly created smart contracts contain 80 lines of code on average. We chose an average of 80 lines of code based on this 1st round of pilot, which was sufficient to contain common vulnerabilities in the ERC20 form. We then embedded at least two common smart contract security vulnerabilities in each contract.

\noindent {\bf Second round pilot.} To test the updated study materials, we had a second round of pilot with another two participants who were blockchain researchers/developers.  
The estimated time for the interview session was 30 minutes and 25 minutes for code review tasks and 5 minutes for exit interview. Specifically, we gave them 20 minutes to review the code (each participant was assigned only one task). They were asked to: a) review the contract code; b) update the code if necessary; c) can search for any resources online during the task. In addition, we gave them 5 minutes to discuss the area of improvement and their rationale after the task. 

% \begin{comment}
% We updated the code review time and protocol which are following: \\ 
% \begin{itemize}
% \item sharing screen so that we can observe  how they are solving or reviewing the task
% \item  20 minutes for task (each participants were assigned with only one task). They were asked to: a) review the contract code; b) updated code if necessary; c) can search for any resources online during task.
% \item Discussing the area of improvement and rationale after the task. (5 minutes)
% \end{itemize}

% \end{comment}
%
%
They made suggestions on the interview questions. For instance, they suggested that we add questions to understand what role(s) the participant played in their smart contract projects. 
%
%Participant D mentioned \textit{``I think it was great. I definitely felt a bit of time pressure, but not too much. It's sometimes hard to waste a lot of time just getting stuck on a little thing or going to load the right documentation. So I could easily imagine some developers who might actually be fully following this and ready to catch some bugs.''}. He also mentioned that \textit{``allowing more time if they thought it was necessary or even doing something like if you need to spend some time loading up documentation, maybe not count that against the time''}.  During the exit interview, Participant D also suggest to include question that cover different level of smart contract developer based on their role. He mentioned \textit{``you're able to get like some smart contract developers from actual blockchain companies, they would definitely be able to respond to some question differently. I kind of imagine that there's a larger population of people with some experience with Solidity in demos, but maybe that haven't written production code.  I mean, maybe there's a way to modify the questions so that you get useful results. Maybe there is someone who's like used a Defi project and interacted with a contract, but not written a contract that's been deployed in production.''} .  
%
Therefore, we added questions to learn: a) what types of role they play in smart contract projects, and b) if they have any experience in deploying smart contracts in a production system or mainnet/testnet. %We presume that this can help us to ask more concentrated follow-up questions to understand their security practices in their consecutive role.
They thought the smart contracts for review were good, but suggested making them ready to go compile from GitHub, which we did in the final study.

%Next, we will present our final study procedures in detail. 

%\subsection{Language selection}
%We selected to use Solidity as the programming language for our experiment of smart contract code review, as it is widely used across many communities for its' usable syntax \cite{parizi2018smart}. Additionally, it has some others support for all kinds of security-related APIs. Another main reason: we are dominantly referring Ethereum Blockchain and solidity is a new programming language designed to create Ethereum-based smart contracts. Therefore, we used solidity for creating smart contracts for this study. As a bonus, Solidity is easier to read and write, is widely used among both beginners and experienced smart contract programmers. Therefore, we reasoned that we would be able to recruit sufficient solidity smart contract developers for our study.
% }

\subsection{Figure: Factors considered in Development and Tools/Programming Languages}
\begin{figure*}[!t]
\centering
\begin{minipage}[b]{0.45\linewidth}
 \includegraphics[width=0.98\columnwidth]{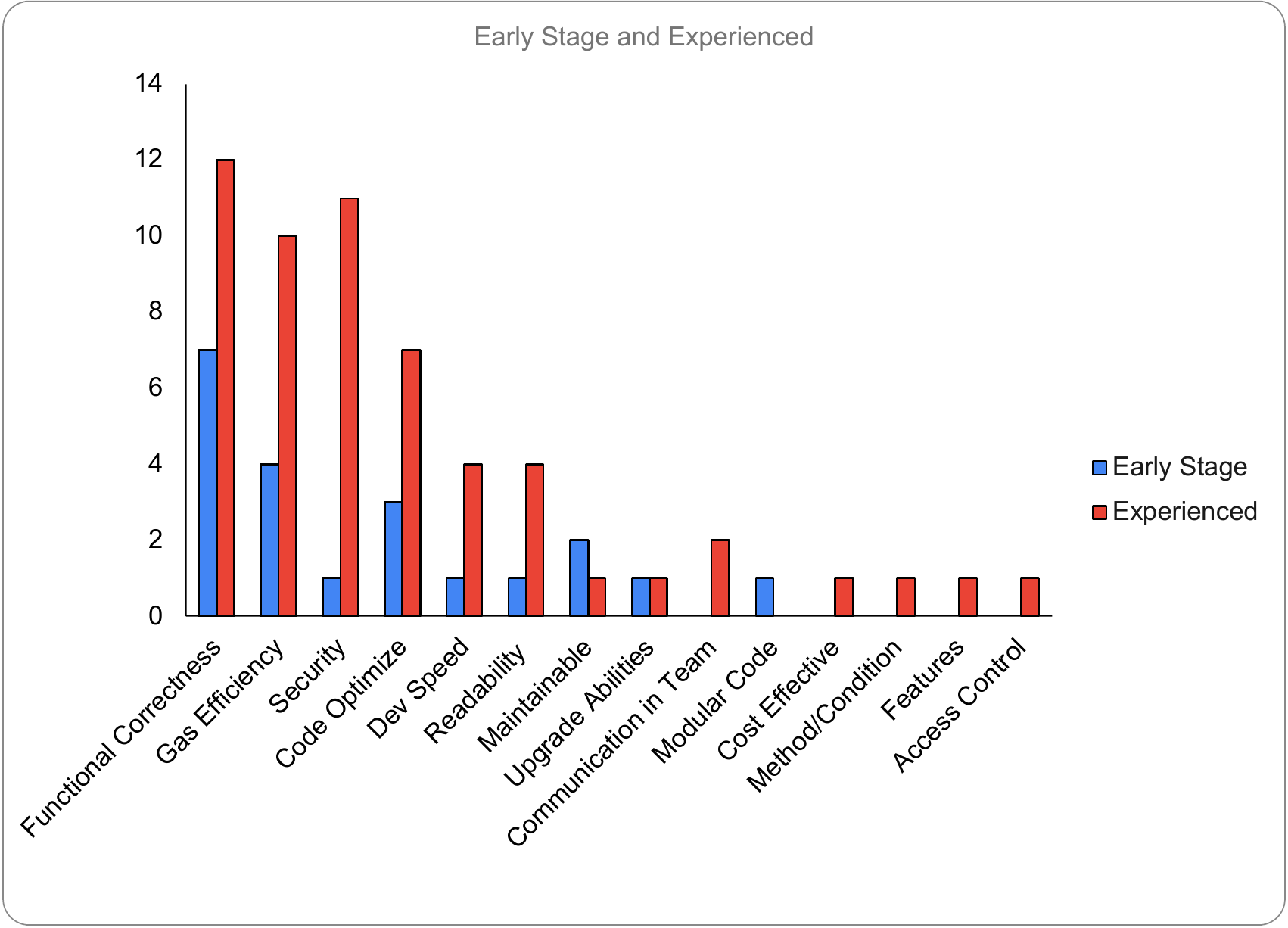}
  \caption{Factors considered in smart contract development: Early stage Vs Experienced Developers}
  \label{fig:pic-factor-comp} 
\end{minipage}
\quad
\begin{minipage}[b]{0.45\linewidth}
 \includegraphics[width=0.98\columnwidth]{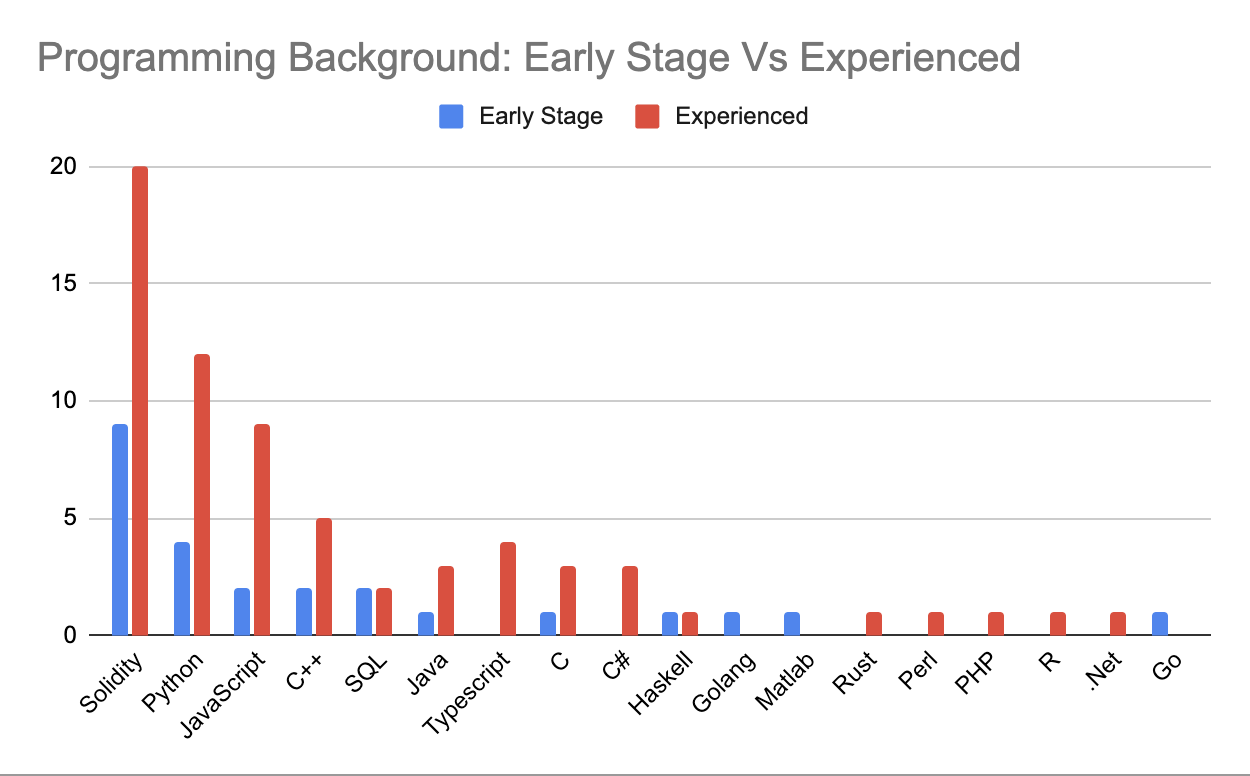}
  \caption{Participant's programming background: Early stage Vs Experienced Developers}
   \label{fig:prog-background}
\end{minipage}
\end{figure*}

We include figures representing different factors considered by early stage and experienced developers (Figure~\ref{fig:pic-factor-comp}), %tools used by participants (early stage and experienced developers) for smart contract development (Figure~\ref{fig:tools}), 
and their programming language background (Figure~\ref{fig:prog-background}). 
\label{app:Extra-Figures}

% \begin{figure}[H]
%   \includegraphics[width=0.7\columnwidth]{factor-v3.png}
%   \caption{Priorities or factors considered in smart contract development}
%   \label{fig:pic-factor}
% \end{figure}

% \begin{figure}
%   \includegraphics[width=0.98\columnwidth]{factor-early-exp-v2.pdf}
%   \caption{Factors considered in smart contract development: Early stage Vs Experienced Developers}
%   \label{fig:pic-factor-comp}
% \end{figure}

% \begin{comment}
% \begin{figure}[H]
%   \includegraphics[width=0.7\columnwidth]{tools-v2.png}
%   \caption{Tools used mentioned by early stage and experienced developers during interview}
%   \label{fig:tools}
% \end{figure}

% \begin{figure}[H]
%   \includegraphics[width=0.7\columnwidth]{tool-early-exp.png}
%   \caption{Tools used mentioned by participants during interview:Early stage and Experienced Developers}
%   \label{fig:tools}
% \end{figure}

% \end{comment}

% \begin{figure}[H]
%   \includegraphics[width=0.8\columnwidth]{prog-v2.png}
%   \caption{Participant's programming background}
%   \label{fig:prog-background}
% \end{figure}

% \begin{figure}
%   \includegraphics[width=0.98\columnwidth]{prog-early-exp.png}
%   \caption{Participant's programming background: Early stage Vs Experienced Developers}
%   \label{fig:prog-background}
% \end{figure}

\subsection{Figure: Task Flowcharts of Participants}
This section includes flowcharts of code review process which were most common among participants. 
\label{app:task-flow and time}
\begin{figure}
  \includegraphics[width=0.98\columnwidth]{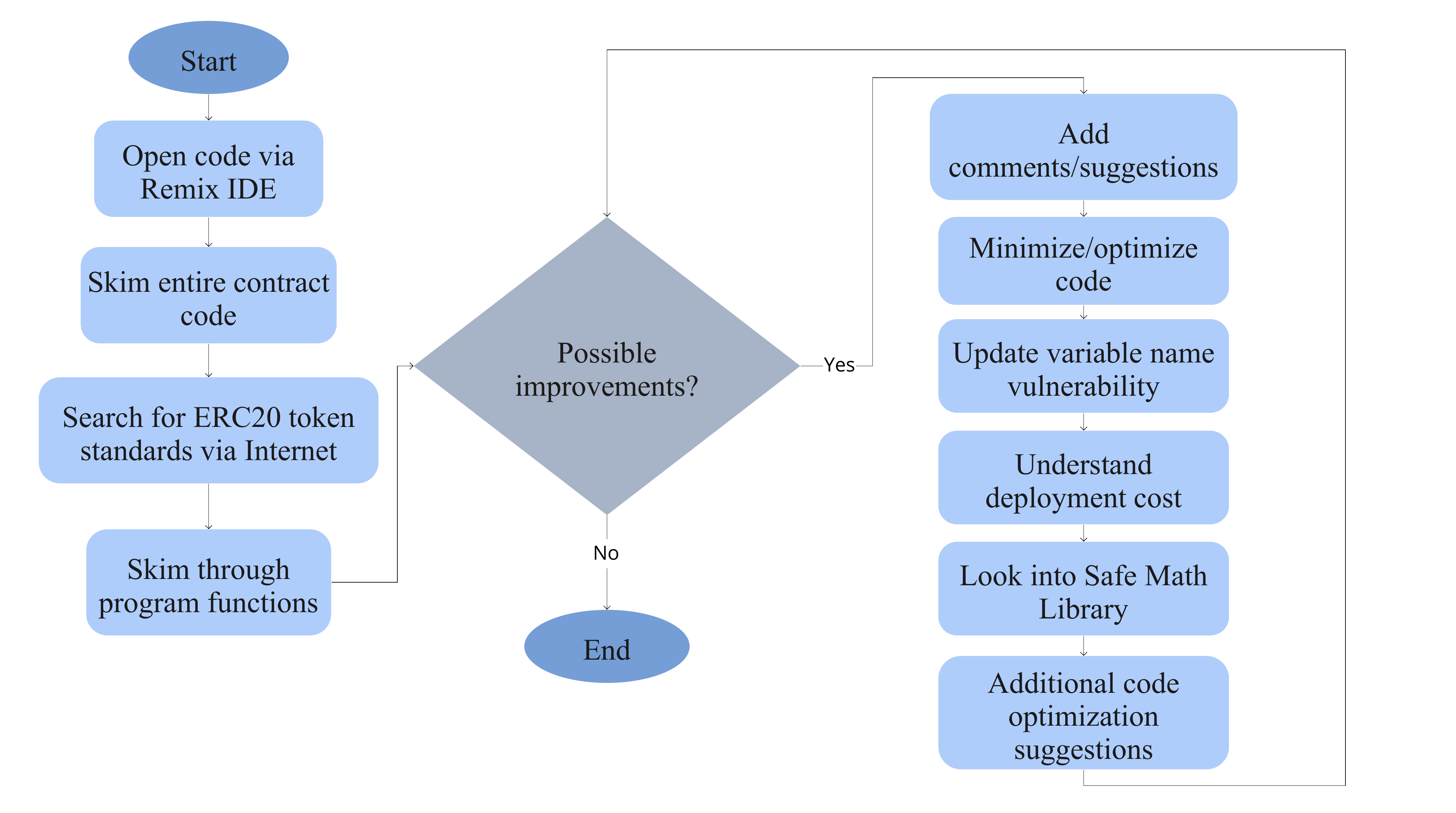}
  \caption{Code Review Task Flowchart for P22 who manually reviewed smart contract for Task 2 and failed to identify any vulnerabilities}
  \label{fig:flow-dia-3}
\end{figure}

% \begin{figure}[H]
%   \includegraphics[width=0.7\columnwidth]{P1.pdf}
%   \caption{Code Review Task Flowchart for P1.}
%   \label{fig:flow-dia-3}
% \end{figure}

% \subsection{Code Snippets Containing Improvement Suggestions}
% \label{app:Extra-code snippets}
% This section includes examples of modified code snippets by participants who were able to identify security vulnerabilities in given code review tasks, and provide modified mode that could fix the vulnerabilities.

\subsection{Code Review Improvement Suggestions: Task 1 and 2}
\label{app:Extra-code snippets}
This section includes examples of modified code snippets by participants who were able to identify security vulnerabilities.

% \begin{comment}

% \begin{figure}[!t]
% \begin{minted}[breaklines,frame=single,fontsize=\small]{python}
%  function batchTransfer(address[] _receivers, uint256 _value) public whenNotPaused returns (bool) {
%     uint cnt = _receivers.length;
%     require(cnt > 0 && cnt <= 20);
%     uint256 amount = mul(uint256(cnt), _value);
    
%     // below is possible taken care of by super.transfer 
%     require(_value > 0 && balances[msg.sender] >= amount);

%     // balances[msg.sender] = balances[msg.sender].sub(amount);
%     for (uint i = 0; i < cnt; i++) {
%         // checks if receivers are address(0x0) --> revert if true
%         super.transfer( _receivers[i], _value);
%         // balances[_receivers[i]] = balances[_receivers[i]].add(_value);
%         // emit Transfer(msg.sender, _receivers[i], _value);}
%     return true;
%     }
% }
% \end{minted}
% \caption{Part of the contract code P6 reviewed and modified. The first few lines had a overflow vulnerability where local variable is calculated as the product of $cnt$ and $\_value$. By having two $\_receivers$ passed into $batchTransfer$, with that extremely large $\_value$, attacker can overflow amount and make it zero.}
% \label{fig:code-snippet-2}
% \end{figure}

% \end{comment}

% \subsection{Task-2 Code Review Improvement Suggestions}

\begin{figure}
% \begin{minted}[breaklines,frame=single,fontsize=\small]{python}
% \begin{minted}{python}
\begin{lstlisting}[breaklines,frame=single]{python}
  modifier onlyOwner() {
    require(msg.sender == address(owner)) ;
    _; 
  }
\end{lstlisting}
\caption{Security vulnerability identification (access control) and modification by P6 during the task2. Tools used: Remix, manual inspection.}

\label{fig:code-snippet-8}
\end{figure}

\begin{figure}
% \begin{minted}[breaklines,frame=single,fontsize=\small]{python}
% \begin{minted}{python}
\begin{lstlisting}[breaklines,frame=single]{python}
     function transferFrom(address from, address to, uint256 tokens)
     override public returns (bool success) {if (balanceOf[from] >= tokens 
     && allowance[from][msg.sender] >= tokens) {
            // no use of safemath
            // todo: is this safe, with the caveat that total supply of 15e23 + 
            //all ether < type(uint256).max
            balanceOf[to] += tokens;
            balanceOf[from] -= tokens;
            allowance[from][msg.sender] -= tokens;
            emit Transfer(from, to, tokens);
            return true;} else {
            // typically a lot of contracts revert in this case
            // revert('Insufficient allowance'); instead of `return false`
            return false; }}
\end{lstlisting}
\caption{Security vulnerability identification (low-level call) and modification by P10 during the task. Tools used: VS Code, manual inspection.}
\label{fig:code-snippet-3}
\end{figure}

\begin{figure}
% \begin{minted}[breaklines,frame=single,fontsize=\small]{python}
% \begin{minted}{python}
\begin{lstlisting}[breaklines,frame=single]{python}
    function withdraw(uint256 amount) public returns (bool success) {
        // typically a lot of contracts revert in this case
        // revert('Insufficient balance'); instead of `return false`
        if (balanceOf[msg.sender] < amount) return false;
        balanceOf[msg.sender] -= amount; // safe because check above
        totalSupply -= amount;
        (bool success,) = msg.sender.call{value: amount}(amount);
        require(success);
        return true;}
\end{lstlisting}
\caption{Security vulnerability identification (re-entrancy) and modification by P10 during the task. Tool used: Remix.}
\label{fig:code-snippet-4}
\end{figure}

% \begin{figure}[!t]
% \centering
%   \includegraphics[width=1.0\columnwidth]{pic12-2.png}
%   \caption{Security vulnerability identification (re-entranccy) and modification by P12 during the task. Tool used: Remix.}
%   \label{fig:12-2}
% \end{figure}

% \begin{figure}[!t]
% \centering
%   \includegraphics[width=1.0\columnwidth]{pic12-1.png}
%   \caption{Security vulnerability identification (low-level unchecked calls) and modification by P12 during the task. Tools used: VS Code, manual inspection.}
%   \label{fig:12-1}
% \end{figure}

\begin{figure}
% \begin{minted}[breaklines,frame=single,fontsize=\small]{python}
% \begin{minted}{python}
\begin{lstlisting}[breaklines,frame=single]{python}
    function withdraw(uint256 amount) public returns (bool success) {
        if (balances[msg.sender] < amount) return false;
        balances[msg.sender] -= amount;
        _totalSupply -= amount;
        //There is some danger in using send: check
        //https://docs.soliditylang.org/
        if (!msg.sender.send(amount)) {
        //vulnerable to reentrancy attack: 
        //https://docs.soliditylang.org/
            balances[msg.sender] += amount;
            _totalSupply += amount;
            return false;}
        return true;}  }
\end{lstlisting}
\caption{Security vulnerability identification (re-entranccy) and modification by P12 during the task. Tool used: Remix.}
\label{fig:12-2}
\end{figure}

\begin{figure}

% \begin{minted}[breaklines,frame=single,fontsize=\small]{python}
% \begin{minted}{python}
\begin{lstlisting}[breaklines,frame=single]{python}
    function transferFrom(address from, address to, uint256 tokens) 
    override public returns (bool success) {
        //modification: adding require here
        require(from!=address(0), "Empty address is not allowed");
        require(to!=address(0), "Empty address is not allowed");
        if (balances[from] >= tokens && allowed[from][msg.sender] >= tokens && tokens> 0) {
            balances[to] += tokens;
            balances[from] -= tokens;
            allowed[from][msg.sender] -= tokens;
            emit Transfer(from, to, tokens);
            return true;
        } else { return false; }}
\end{lstlisting}
\caption{Security vulnerability identification (low-level unchecked calls) and modification by P12 during the task. Tools used: VS Code, manual inspection.}
\label{fig:12-1}
\end{figure}

\begin{figure}
% \begin{minted}[breaklines,frame=single,fontsize=\small]{python}
% \begin{minted}{python}
\begin{lstlisting}[breaklines,frame=single]{python}
 //Might be a vulnerability- something to do with miners reordering approve txs
 //I kinda remember approve funtions having a "set to 0" kinda logi
//look deeper later
        function approve(address spender, uint256 tokens) override public returns (bool success) {
        allowed[msg.sender][spender] = tokens;
        emit Approval(msg.sender, spender, tokens);
        return true;
    }
\end{lstlisting}
\caption{False positive Security vulnerability identification for task1 by P1}
\label{fig:P1-false}
\end{figure}

\label{code_review}

\begin{table*}[htbp]
    \centering
    \caption{Summary of participants' task performance in the code review task where success rate (i.e., Rate) denotes the percentage of participants who correctly identified a particular vulnerability; 0 : the participant did not identify the vulnerability; 1: the participant identified the vulnerability.
    The middle section of the table shows the early-stage (< 1 year of experience) Solidity developer participants, whereas the bottom section of the table shows the experienced (2-5 years of experience) Solidity developer participants. Every participant either did Task 1 or Task 2. Task 1 includes two  vulnerabilities: reentrancy and unchecked low-level call. Task 2 includes two  vulnerabilities: integer overflow and improper access control. We also show the review method each participant used during the task. Most participants did manual inspection of the code using an IDE or a test editor such as Remix, VS Code, Sublime, IntelliJ, or simply viewing the code on GitHub. Remix and Hardhat are development/testing frameworks for smart contracts. Three participants (P3, P15, P20) used Remix with a security plugin (sp), which does Solidity static analysis. Slither is a static analysis tool and Oyente is an analysis tool based on symbolic execution.}  
    \small
    \begin{tabular}%{*{8}{p{.095\linewidth}}}%
    {l l l l l l l l }
    \hline
   ID &  {\bf Reentrancy} & {\bf \shortstack{Low-Level\\ Call}} &{\bf \shortstack{Review\\Method}}  &ID & {\bf Overflow}  & {\bf \shortstack{Access\\ Control}} &{\bf\shortstack{Review\\Method}} \\
   \hline
  (Task 1) & Success & Success &  & (Task 2) & Success & Success & \\
    &  Rate: 53.3\%& Rate: 40\%&  &  & Rate: 28.6\% & Rate: 42.9\%& \\
 \hline
\textbf{Early} & Rate: 25\% & Rate: 0\% &  & & Rate: 16.7\% & Rate: 16.7\%\\
%\hline
P4  & 0 & 0   & Manual & P1  &0 &0 &Manual\\
P5  &  0 &0   & Manual & P3  &0  &0  & Remix (sp)\\
P7  & 0 &0   & Manual  & P9  &0  &0 & Manual  \\
P20  & 1& 0  & Remix (sp) & P24  & 0   &0 & Manual\\ 
& & & & P16 & 0 &0 &Manual\\
& & & & P28  & 1 &1& Manual \\
\hline
\textbf{Exp.} & Rate: 63.6\% & Rate: 54.5\% & & & Rate: 37.5\% & Rate: 62.5\% \\
%\hline
P2 & 0  & 0 & Manual &P8  &1 &1 &Manual\\
P10  &1 &1 & Manual & P6 &1& 1& Manual\\
P11  & 0  &0 & Manual &P13  &0  &0 &Manual\\
P12  &1 &1 & Manual &P15  & 0  & 1 &Remix (sp)\\
P14  & 1 &1  & Hardhat+Slither & P17  & 1& 0 & Manual \\
P18  &1 &0  & Manual & P21  & 0  & 1&Remix+Oyente\\
P19  & 1 &0   & Manual &  P22  & 0   &0 & Manual\\
P23  &0  &0  & Manual &  P26  & 0   &1& Manual\\ 
P25  &1 &1 & Manual+Remix (sp) \\ 
P27  & 1  &1& Manual+MythX \\
P29  & 0   &1& Manual+Remix (sp) \\

   \hline
   
    \end{tabular}
    %\label{Tab:task-per}
    % \begin{tablenotes}
   
    % \end{tablenotes}

%\label{table:app_func}
\label{Tab:tab-per}
  \end{table*}

 \begin{table*}[t!]
    \centering
    \caption{Average time for identifying each vulnerability by early stage and experienced developers. $avg_t$: Average Time} 
  
    \small
    \begin{tabular}%{*{8}{p{.095\linewidth}}}%
    {l l l | l l l }
   \hline
   ID &  {\bf Reentrancy} & {\bf Low-Level Call} &ID & {\bf Overflow}  & {\bf Access Control} \\
 
  \hline
\textbf{Early Stage} &  & &  & & \\

P20  & 18.18 &  & P28 & 7.07& 8.31 \\
  & $avg_t$: \textbf{18.18} &   &$avg_t$: \textbf{7.07} & & $avg_t$: \textbf{8.31}\\
\hline
\textbf{Experienced} & &  & & &   \\

P10 & 5.08 & 8.08 &  P6  &9.08 &14.15 \\
P12  &11.98 & 8.16 &   P8 &9.17& 14.01\\
P14  & 12.44 & 10.12 &  P15  & & 11.45  \\
P18  & 13.07 &  &   P17  & 9.13 &   \\
P19  & 13.50 & &    P21  &  & 14.54\\
P25  & 6.23 &10.53 &    P26  &  & 15.47\\
P27  & 12.13 &11.56 &   \\
P29  &  &10.37 &   \\
 & $avg_t$: \textbf{10.6} & $avg_t$: \textbf{10.4} &    & $avg_t$: \textbf{10.1} & $avg_t$:  \textbf{14.4} \\

 \hline
   
    \end{tabular}
    %\label{Tab:task-per}
    % \begin{tablenotes}
   
    % \end{tablenotes}

%\label{table:app_func}
\label{Tab:tab-time}
  \end{table*}

\end{document}